\pdfoutput=1

\documentclass[11pt,twoside,a4paper,cmspaper,final,collab]{cms-tdr}
\def\svnVersion{166737}\def\cmsCernNo{2012-093}\def\cmsCernDate{\today}\def\cmsMessage{Submitted to Physics Letters B}

\begin{document}\cmsNoteHeader{EXO-11-022}

\hyphenation{had-ron-i-za-tion}
\hyphenation{cal-or-i-me-ter}
\hyphenation{de-vices}

\RCS$Revision: 125285 $
\RCS$HeadURL: svn+ssh://svn.cern.ch/reps/tdr2/papers/EXO-11-022/trunk/EXO-11-022.tex $
\RCS$Id: EXO-11-022.tex 125285 2012-05-29 18:37:20Z alverson $
\ifthenelse{\boolean{cms@external}}{\providecommand{\cmsLeft}{Top}}{\providecommand{\cmsLeft}{Left}}
\ifthenelse{\boolean{cms@external}}{\providecommand{\cmsRight}{Bottom}}{\providecommand{\cmsRight}{Right}}
\cmsNoteHeader{EXO-11-022} 
\title{Search for heavy long-lived charged particles in pp collisions at $\sqrt{s}=7\TeV$}

\date{\today}
\newcommand{\stau}{\ensuremath{\widetilde{\tau}_1}\xspace}
\newcommand{\smu}{\ensuremath{\widetilde{\mu}_1}\xspace}
\newcommand{\sel}{\ensuremath{\widetilde{\mathrm{e}}_1}\xspace}
\newcommand{\stp}{\ensuremath{\widetilde{\mathrm{t}}_1}\xspace}
\newcommand{\gluino}{\ensuremath{\mathrm{g}}\xspace}
\newcommand{\ETM}{\ET}
\newcommand{\sto}{\ensuremath{\widetilde{\mathrm{t}}_1}\xspace}
\newcommand{\ias}{\ensuremath{I_\mathrm{as}}}
\newcommand{\ih}{\ensuremath{I_\mathrm{h}}}
\newcommand{\gsim}{\gtrsim}%
\newcommand{\hyperk}{\ensuremath{\widetilde {\mathrm{K}}}\xspace}
\newcommand{\hyperkbar}{\ensuremath{\overline{\widetilde {\mathrm{K}}}}\xspace}
\newcommand{\hyperpi}{\ensuremath{\widetilde{\pi}}\xspace}
\newcommand{\hyperpibar}{\ensuremath{\overline{\widetilde{\pi}}}\xspace}
\newcommand{\hyperrho}{\ensuremath{\widetilde{\rho}}\xspace}
\providecommand{\rd}{\ensuremath{\mathrm{d}}} 
\newcommand{\dedx}{\ensuremath{\rd E/\rd x}\xspace}

\abstract{
The result of a search for heavy long-lived charged
particles produced in pp collisions at $\sqrt{s} = 7\TeV$ at the LHC is
described.
The data sample has been collected using the CMS detector and corresponds to
an integrated luminosity of 5.0\fbinv.
The inner tracking detectors are used to define a sample of events
containing tracks with high momentum
and high ionization energy loss.
A second sample of events, which have high-momentum tracks
satisfying muon identification requirements in addition to meeting
high-ionization and long time-of-flight requirements, is analyzed
independently.
In both cases, the results are consistent with
the expected background estimated from data.
The results are used to establish cross section limits as a function of mass
within the context of models with long-lived gluinos, scalar
top quarks and scalar taus. Cross section limits on hyper-meson
particles, containing new elementary long-lived hyper-quarks
predicted by a vector-like confinement model, are also presented.  Lower limits
at 95\% confidence level on the mass of gluinos (scalar
top quarks) are found to be 1098 (737)\GeVcc.  A limit of 928
(626)\GeVcc is set for a gluino (scalar top quark)
that hadronizes into a neutral bound state before reaching the muon
detectors. The lower mass limit for a pair produced scalar tau is found to be
223\GeVcc.
Mass limits for a hyper-kaon are placed at 484, 602, and 747\GeVcc for hyper-$\rho$ masses of 800, 1200, and 1600\GeVcc,
respectively.
}

\hypersetup{%
pdfauthor={CMS Collaboration},%
pdftitle={Search for heavy long-lived charged particles in pp collisions at sqrt(s)=7 TeV},%
pdfsubject={CMS},%
pdfkeywords={CMS, physics}}

\maketitle 

\section{Introduction}

Various extensions to the standard model (SM) of particle physics
allow for the possibility that as-yet-undiscovered massive
(${\gtrsim}100$\GeVcc) elementary particles could be long-lived
with lifetime greater than ${\sim} 1$\unit{ns} as a result of a new
conserved quantum number, a kinematic constraint or a weak
coupling~\cite{Drees:1990yw, Fairbairn:2006gg, Bauer:2009cc}.
Such particles, where they are electrically charged, are referred to
as Heavy Stable Charged Particles (HSCP) in this article. Because of
their high mass, a significant fraction of the HSCPs that could be produced at the
Large Hadron Collider (LHC) are expected to be detectable as high
momentum  ($p$) tracks with an anomalously large rate of energy loss
through ionization ($\rd E/\rd x$) and an anomalously long time-of-flight (TOF).

Previous collider searches for HSCPs have been performed at
LEP~\cite{Barate:1997dr, Abreu:2000tn,Achard:2001qw, Abbiendi:2003yd},
HERA~\cite{Aktas:2004pq}, the
Tevatron~\cite{Abe:1989es, Abe:1992vr, Acosta:2002ju, Abazov:2007ht,
Abazov:2008qu, Aaltonen:2009kea, PhysRevLett.108.121802},  and the
LHC~\cite{PhysRevLett.106.011801,Khachatryan:2011ts,Aad:2011mb,Aad:2011yf,Aad:2011hz,Aad:2012zn}.
HSCPs are expected to reach the outer muon systems of the
collider detectors even if they are strongly interacting. In that case it
is expected that a bound state ($R$-hadron) is formed in the process
of hadronization~\cite{Farrar:1978xj,Fayet:1983ye,Farrar:1984gk} and
that the energy loss occurs primarily through low momentum transfer
interactions~\cite{Drees:1990yw, Baer:1998pg,
Mafi:1999dg,Kraan:2004tz}, allowing the $R$-hadron to traverse an amount
of material typical of the calorimeter of a collider
experiment. However, the nuclear interactions experienced in matter by
an $R$-hadron may lead to charge exchange.
A recent study~\cite{Mackeprang:2009ad}
of the modeling of nuclear interactions of HSCPs
favours a scenario in which the majority
of the $R$-hadrons containing a gluino, \PSg\ (the supersymmetric
partner of the gluon), or a bottom squark would emerge
neutral in the muon detectors.
Given the large uncertainties in the nuclear interactions
experienced by $R$-hadrons, experimental strategies that do not rely
on a muon-like behavior of HSCPs are important.
Two strategies already employed are to search for very slow
($\beta \lesssim 0.4$) $R$-hadrons brought to rest in the
detector~\cite{Abazov:2007ht,PhysRevLett.106.011801,Aad:2012zn},
and to search using only inner tracker and/or calorimeter
information~\cite{Khachatryan:2011ts,Aad:2011mb,Aad:2011yf}.

In this article we present a search for HSCPs
produced in pp collisions at $\sqrt{s} = 7$ TeV at the LHC,
recorded using the Compact Muon Solenoid (CMS)
detector~\cite{Chatrchyan:2008zzk}. The search is based on
a data sample collected in 2011, corresponding to an
integrated luminosity of $5.0$ fb$^{-1}$.
Events were collected using either of two triggers, one based on
muon transverse momentum (\pt) and the other based on the missing
transverse energy (\MET) in the event. This event sample was then
used for two separate selections. In the first, the HSCP candidates
were defined as tracks reconstructed in the inner tracker detector
with large \dedx and high \pt. In the second, the tracks were also
requested to be associated with identified muons that had a long time-of-flight
as measured by the muon detectors. The first selection is largely
insensitive to the uncertainties in modeling the $R$-hadron nuclear interactions.
For both selections, the mass of the candidate was calculated from
the measured values of $p$ and \dedx.
This analysis extends our previously published
result~\cite{Khachatryan:2011ts}
through the use of a larger dataset, muon TOF information, and track
isolation requirements. This new analysis also probes
additional signal models.

\section{Signal benchmarks \label{sec:signalmc}}

The results of this search have been interpreted within the context of
several theoretical models. Supersymmetric
models~\cite{Martin:1997ns, Kane:2010zza} can in some cases allow for HSCP
candidates in the form of gluinos, scalar top quarks (stops, \stone,
the supersymmetric partner of the top quark), and scalar taus (stau,
\stau, the supersymmetric partner of the $\tau$).
We also consider a new model that postulates a QCD-like confinement
force between new elementary particles  (hyper-quarks, $\widetilde{
q}$)~\cite{Kilic:2009mi} and allows for long-lived hyper-mesons.

In order to study the uncertainties
related to the underlying production processes, samples were produced
with three different multiparton interaction (MPI)
models~\cite{Skands:2010ak}: DW with CTEQ5L parton distribution
functions (PDF)~\cite{Lai:1999wy} (used in~\cite{Aad:2011yf}), D6T with
CTEQ6L1 PDF~\cite{Pumplin:2002vw} (used in~\cite{Khachatryan:2011ts}),
and Z2 with CTEQ6L1 PDF. The latter model features a harder
initial-state radiation spectrum. The final results of this analysis are
obtained with samples using the D6T MPI, which yield the most conservative
signal selection efficiency of the three choices.

Gluino and stop production were modelled as pair production over
the particle mass range 130--1200\GeVcc using \PYTHIA
v6.422~\cite{Sjostrand:2006za}.
For \PSg\ production, we set the squark masses to very high
values ($>$7 TeV) to reflect the scenario of split
supersymmetry~\cite{ArkaniHamed:2004fb, Giudice:2004tc}.
The fraction $f$ of produced \PSg\  hadronizing
into a \PSg-gluon state ($R$-gluonball) is an unknown
parameter of the hadronization model and affects the fraction of
$R$-hadrons that are produced as neutral particles. As
in~\cite{Khachatryan:2011ts},
results were obtained for two different values of $f$, 0.1 and
0.5. Unless specified otherwise, the value $f=0.1$ is assumed.
The interactions of the HSCPs using the CMS apparatus and the detector
response were simulated in detail with the \GEANTfour
v9.2~\cite{Agostinelli2003250, Allison:2006ve} toolkit.
Two scenarios for $R$-hadron strong interactions with
matter were considered: the first follows the model defined
in~\cite{Kraan:2004tz, Mackeprang:2006gx}, while the second is one of
complete charge suppression, where any nuclear interaction of
the $R$-hadron causes it to become neutral.  In the second
scenario, effectively all $R$-hadrons are neutral by the time
they enter the muon system.

The  minimal gauge-mediated supersymmetry breaking
(GMSB) model~\cite{Giudice:1998bp} was used to describe \stau\
production, which can proceed either via direct pair production
or via production of heavier
supersymmetric particles that
decay to one or more \stau.
The latter process has a larger cross section than direct
production.
Two benchmark points on the Snowmass Points and Slopes
line 7~\cite{Allanach:2002nj} have been considered. They correspond to $N=3$
chiral SU(5) multiplets added to the theory at the
scales $F = 100$ and 160
TeV~\cite{Giudice:1998bp} respectively, and an
effective supersymmetry-breaking scale $\Lambda = 50$ and $80$\TeV
respectively. Both points have the ratio of neutral Higgs field vacuum
expectation values $\tan{\beta}=10$, a positive sign for the
Higgs-Higgsino mass parameter ($\sgn (\mu)=1$), and the ratio of the
gravitino mass to the value it would have if the only supersymmetry
breaking scale were that in the messenger sector, $c_\text{grav}=10^4$.
The particle mass spectrum and the decay table were produced
with the program \ISASUGRA version 7.69~\cite{Paige:2003mg}. The
resulting \stau\ masses are 156 and 247\GeVcc, and
the squark and gluino masses are about 1.1 and 1.7\TeVcc, respectively.
Additional mass points in the range 100 to 500\GeVcc were obtained by
varying the $\Lambda$ parameter in the range 31 to 100\TeV and keeping
the ratio of $F$ to $\Lambda$ equal to 2.
In addition, direct \stau\ pair production samples were generated separately.

As mentioned above, another HSCP benchmark considered in this paper is a
vector-like confinement model that postulates a QCD-like confinement
force between new elementary particles (hyper-quarks, $\widetilde{q}$)~\cite{Kilic:2009mi}. The hyper-quarks can be confined into SM
hadron-like hyper-mesons such as hyper-$\pi$ (\hyperpi), hyper-K
(\hyperk), or hyper-$\rho$ (\hyperrho).
We assume a simplified model (similar to that in Section 4.2 of
Ref.~\cite{Kilic:2009mi}) with \hyperpi\ or
\hyperk\ pair production via either the Drell--Yan process (\hyperk
\hyperkbar) or via production of a resonant \hyperrho\ (\hyperrho $\to$
\hyperk\hyperkbar) analogous to QCD $\rho$ meson production~\cite{Kilic:2010et}.
The \hyperrho\ mixes only with the
electroweak gauge bosons, and therefore is not produced strongly.
In this model, the \hyperpi\ is short-lived and decays to
SM gauge bosons (e.g. \hyperpi\ $\to W^\pm\gamma$,
\hyperpi $\to W^\pm Z$).  Its production processes are
not included in the simulation.
However, the \hyperk\ is long-lived compared to the detector size and
constitutes an HSCP candidate. In the considered model, the \hyperk,
like the \stau, would only interact via the electroweak force.
The mix of resonant
and Drell-Yan
production results in different
kinematics, compared to the GMSB \stau\ model.
For \hyperk\ mass values much less than half the \hyperrho\ mass, the HSCP
receives a significant boost from the resonance
decay~\cite{Chen:2009gu}, while near threshold the
\hyperk\ is slow enough that the acceptance drops dramatically.
Parton-level events were generated with
\CALCHEP v2.5.4~\cite{Pukhov:2004ca} and passed to
\PYTHIA for hadronization and simulation of the underlying event.
The masses of the \hyperk, \hyperpi, and \hyperrho\ are treated as
free parameters  in the model, affecting in particular the \hyperrho\ width.
We used a fixed \hyperpi\ mass of 600\GeVcc and
\hyperk\ masses in the range 100 to 900\GeVcc, for \hyperrho\ masses
of 800, 1200 and 1600\GeVcc.

In all simulated samples, the primary collision event was
overlaid with simulated minimum-bias events to
reproduce the distribution of the number of inelastic collisions per
bunch crossing (pile-up) observed in data.

\section{The CMS detector}
A detailed description of the CMS detector can be
found elsewhere~\cite{Chatrchyan:2008zzk}.
The central feature of the CMS apparatus is a
superconducting solenoid of 6\unit{m} internal diameter.
Within the field volume are the silicon pixel and strip
inner tracking detectors, the crystal electromagnetic calorimeter
(ECAL), and the brass/scintillator hadron calorimeter (HCAL). Muons
are measured in gas-ionization detectors embedded in the steel return
yoke. In addition to the barrel and endcap detectors, CMS has
extensive forward calorimetry.
CMS uses a right-handed coordinate system, with the origin at the
nominal interaction point, the $x$ axis pointing to the center of the
LHC ring, the $y$ axis pointing up (perpendicular to the LHC plane), and
the $z$ axis along the counterclockwise-beam direction. The polar
angle, $\theta$, is measured from the positive $z$ axis and the
azimuthal angle, $\phi$, is measured in the $x$-$y$ plane.
The muons are measured in the pseudorapidity ($\eta
\equiv -\ln \tan (\theta/2)$) range
$|\eta|< 2.4$, with detection planes made using three technologies:
drift tubes (DT), cathode strip chambers (CSC), and
resistive plate chambers (RPC). The DT and CSC detectors are installed in
the barrel at $|\eta| < 1.2$ and in the endcaps at $0.9 < |\eta| <
2.4$, respectively, whereas RPCs cover the range $|\eta| < 1.6$.
The inner tracker measures charged particle trajectories within the
pseudorapidity range $|\eta| < 2.5$. It consists of 1440 silicon pixel
modules and 15\,148 silicon strip modules. The \pt\ resolution for
tracks measured in the central (forward) region of the silicon tracker
is 1\% (2\%) for \pt\ values up to 50\GeVc and degrades to 10\%
(20\%) at \pt\ values of 1\TeVc.
The CMS trigger
consists of a two-stage system.  The first level (L1) of
the CMS trigger system, composed of custom hardware processors,
uses information from the calorimeters and muon detectors to select
a subset of the events. The High Level
Trigger (HLT) processor farm further decreases the event rate from
around 100\unit{kHz} to around 300\unit{Hz}, before data storage.

\subsection{\texorpdfstring{\dedx}{dE/dx} Measurement}

The \dedx measurement for a candidate track was performed using the charge
information contained in the track measurements provided by the
silicon strip and pixel detectors.
A silicon strip or pixel measurement consists of a cluster of adjacent
strips or pixels with a charge above threshold.
These clusters form the basis for \dedx measurements in this analysis.
For \dedx measurement purposes, a `cleaning procedure' was applied to
the clusters found in the silicon strip detectors. This selection is
intended to reduce anomalous ionization contributions due to overlapping
tracks, nuclear interactions, and hard  $\delta$-rays in the silicon
strip detectors. Genuine single particles release charge
primarily within one or two neighbouring strips.
Other strips generally carry only a fraction (to a first approximation
equal to $10^{-n}$, where $n$ is the distance in units of strips) of the
total cluster charge from capacitive coupling and cross-talk
effects~\cite{:2009dia}. Measurements displaying multiple
charge maxima or more than two adjacent strips containing comparable charge
were therefore not used in the
\dedx calculations.
This cleaning procedure, which discards on average about 20\% of the
track measurements, rejects background at high \dedx without
a significant impact on the signal acceptance.

As in~\cite{Khachatryan:2011ts}, a modified version of the
Smirnov--Cramer--von Mises~\cite{Eadie, James} discriminant was used for
estimating the degree of compatibility of the observed charge
measurements with those expected for particles close to the
minimum of ionization:
\begin{equation}
\ias = \frac{3}{J} \times \left\{
\frac{1}{12J} + \sum_{i=1}^J
\left[
P_i \times \left( P_i - \frac{2i-1}{2J} \right)^2 \right] \right\},
\label{eq:ias}
\end{equation}
where $J$ is the number of track measurements in the silicon-strip detectors,
$P_i$ is the probability for a particle close to
the minimum of ionization to produce a charge smaller or equal to that of the $i$--th measurement
for the observed path length in the detector, and the sum is over the track
measurements ordered in terms of increasing $P_i$.
The charge probability density functions used to calculate $P_i$ were
obtained using reconstructed tracks with $\pt>5$\GeVc in events collected with a
minimum bias trigger.
As in~\cite{Khachatryan:2011ts}, the most probable value of the
particle \dedx was determined using a harmonic estimator \ih:
\begin{equation}
\ih= \biggl( \frac{1}{N} \sum_{i=1}^N (c_{i})^{-2} \biggr)^{-1/2},
\label{eq:HarmonicEstimator}
\end{equation}
where $N$ is the total number of track measurements in the
pixel and silicon-strip detectors and $c_{i}$ is the charge per unit
path length
of the $i$-th measurement.
As implied above,
the \ih\ estimator was computed using both silicon strip and pixel
measurements, whereas the \ias\ estimator was based only on silicon strip measurements.
As in~\cite{Khachatryan:2011ts}, the mass measurement was based on the
formula:
\begin{equation}
\ih= K\cfrac{m^2}{p^2}+C ,
\label{eq:MassFromHarmonicEstimator}
\end{equation}
where the empirical parameters $K=2.559 \pm 0.001\MeV \unit{cm}^{-1}\,c^2$
and $C=2.772 \pm 0.001\MeV\unit{cm}^{-1}$ were determined from data using a
sample of low-momentum protons~\cite{Khachatryan:2010pw}.
Equation~(\ref{eq:MassFromHarmonicEstimator}) reproduces the Bethe-Bloch
formula~\cite{Nakamura:2010zzi} with an accuracy of better than 1\% in
the range $0.4 < \beta < 0.9$, which corresponds to $1.1 <
(\dedx)/(\dedx)_\mathrm{MIP} < 4.0$, where $(\dedx)_\mathrm{MIP}$ is the ionization
energy loss rate of a particle at the minimum of ionization.
Equation~(\ref{eq:MassFromHarmonicEstimator}) implicitly
assumes that the HSCP candidates have unit charge. For HSCPs with masses
above 100\GeVcc, the mass resolution is expected to degrade
gradually with increasing mass. This effect is due to the deterioration of
the momentum resolution and to the limit on the maximum charge that
can be measured by the silicon strip tracker analogue-to-digital
converter modules, which also affects
the mass scale.
These effects are modeled in the simulation. For an HSCP with a
mass of 300\GeVcc, the mass resolution and the most probable
reconstructed mass were found to be 16\% and 280\GeVcc, respectively.

For each reconstructed track with momentum $p$ as measured in the inner
tracker, \ias, \ih\ and $m$ were computed using
Eq.~(\ref{eq:ias}), (\ref{eq:HarmonicEstimator}),
and~(\ref{eq:MassFromHarmonicEstimator}).
These values were used in the candidate selection as described in
Sections~\ref{sec:preselection} and~\ref{sec:background}.

\subsection{TOF Measurement}

A major addition to this analysis with respect to that
in Ref.~\cite{Khachatryan:2011ts} is the use of TOF information
from the muon system.
The $\beta$ measurement for a candidate track was performed using time
information provided by the individual DT and CSC track measurements.
The DT system
consists of four layers of muon chambers interleaved with the return
yoke of the solenoid.
The chambers in the three innermost layers contain three
super-layers (SL) each, two of them measuring the track $\phi$
projection and the third measuring the $\theta$ projection. The
chamber in the outermost layer is equipped with just two SLs that
measure the track $\phi$ projection. Each SL is composed of four DT layers.
For TOF measurement purposes, only SLs providing measurements in the
$\phi$ projection were used because their time resolution is a factor
of two better than that of the $\theta$-projection SLs.
The CSC system comprises four layers of chambers at increasing $\abs{z}$ positions.
Each chamber contains six detection layers. All
detection layers were used for TOF measurement purposes.

Both the DTs and CSCs measure the difference ($\delta_t$) between
the particle crossing time and the average time at which a high-momentum muon,
produced at the nominal collision point in the triggered bunch crossing, would
pass through the same portion of the detector.
Measurements from prompt HSCPs would yield a $\delta_t$
greater than zero.
The $\phi$-projection DT measurements within a chamber were fitted with
a straight line. In order to improve the accuracy of
the parameters of the straight line for late tracks, a time shift common to all
measurements within the chamber was introduced as a third free
parameter of the fit. Having four
chambers with eight layers measuring the track $\phi$ projections,
there are up to 32 independent DT $\delta_t$ measurements along a
candidate track.
Each detection layer in a CSC chamber has a nearly orthogonal
layout of anode wires and cathode strips.  The arrival time of the signals
from both the anode wires and cathode strips measures the particle $\delta_t$.
Having four chambers, six detection layers per chamber, and two
$\delta_t$ measurements per layer, there are up to 48 independent CSC
$\delta_t$ measurements along a candidate track.

A single $\delta_t$ measurement can be used to determine the track
$\beta^{-1}$ via the equation:
\begin{equation}
\beta^{-1}= 1+ \frac{c \delta_t}{L}
\label{betatotof}
\end{equation}
where $L$ is the flight distance and $c$ is the speed of light.
The track $\beta^{-1}$ value was calculated as the weighted
average of the  $\beta^{-1}$ measurements associated with the track.
The weight for the $i^{th}$ DT measurement is given by:
\begin{equation}
w_{i} = \frac{(n-2)}{n}\frac{L_{i}^{2}}{\sigma_{DT}^{2}}
\end{equation}
where $n$ is the number of $\phi$ projection measurements found in the
chamber from which the measurement comes
and  $\sigma_{DT}$ is the time resolution of DT
measurements, for which the measured value of 3\unit{ns} is used. The factor
$(n-2)/n$ arises from computing measurement residuals in a
plane and with respect to a straight line resulting from a fit to the same
measurements.
The weight for the $i^{th}$ CSC measurement is given by:
\begin{equation}
w_{i} =\frac{L_{i}^{2}}{\sigma_{i}^{2}}
\end{equation}
where $\sigma_{i}$, the measured time resolution, is 7.0\unit{ns} for cathode strip
measurements and 8.6\unit{ns} for anode wire measurements.

To reduce the impact of outliers, which are mostly observed in the CSC
anode wire measurement distribution, the CSC $\beta^{-1}$
measurement whose difference with the
track averaged $\beta^{-1}$ is largest was discarded if the difference
was greater than three times the estimated uncertainty
($\sigma_{\beta^{-1}}$) in the track averaged $\beta^{-1}$. The track averaged
$\beta^{-1}$ and the associated uncertainty were recomputed without
the excluded measurement and the procedure was iterated until no
further measurements could be discarded.
A Gaussian fit to the core of the distribution of the
$\beta^{-1}$ measurements for the candidates passing the
muon-like track pre-selection defined in
Section~\ref{sec:preselection} yielded a width of approximately 0.06,
independent of the candidate pseudorapidity.

\section{Trigger and data selection \label{sec:preselection}}

Events were selected using a trigger requiring a muon with high
transverse momentum ($\pt> 40\GeVc)$ with $|\eta|< 2.1$,
or a trigger requiring large missing transverse energy ($\MET >150$
\GeV). The latter quantity was computed online using jets reconstructed with
a particle-flow algorithm~\cite{JME-10-003}.
Jet clustering was performed using the
anti-$k_\mathrm{T}$ algorithm~\cite{antikt} with a size parameter of 0.5.
Triggering on $\MET$ allows the recovery of events with HSCPs failing muon
identification or emerging mainly as neutral particles after traversing the
calorimeters.
The L1 muon trigger accepts tracks that produce signals in the
RPC detectors either within the 25\unit{ns} time window corresponding to the
collision bunch crossing, or within the following 25\unit{ns} time window.
This operation mode is particularly suited for detecting late tracks
in the muon system. It was designed to cater for this analysis, and is
tenable as long as collisions are separated by 50\unit{ns} or more, which
was the case for the 2011 LHC running period.
The DT and CSC L1 triggers were
used only for detecting particles produced in the collision bunch crossings.
Track reconstruction in the muon HLT assumes particles
traveling at the speed of light and produced within the triggered
bunch crossing. However, the requirements
on the quality of the muon segments are loose enough to allow
tracks from late particles to be reconstructed with reasonably high
efficiency. Events with pair produced \stau, and with the fastest
\stau\ having $\beta$ as low as 0.6, would be selected by the muon trigger
with 75\% efficiency. The muon trigger efficiency would become less
than 10\% for events where the fastest \stau\ had $\beta \leq 0.45$.

The analysis made use of two offline selections referred to as
`tracker only' and `tracker+TOF'. In the tracker-only selection,
HSCP candidates were defined as individual tracks reconstructed in
the inner tracker with large \dedx and \pt.
In the tracker+TOF selection, the track was additionally required to
be associated with an identified muon with long TOF.
Events selected online with either of the muon or $\MET$ triggers were
used in each of these two offline selections, to maximize the
acceptance for HSCP signals. As described in section~\ref{sec:results}, the
uncertainty in the signal acceptance arising from the uncertainty in
the trigger efficiency is also reduced for some of the signals,
because of the overlap of the two triggers. The tracker+TOF selection
is not a subset of the tracker-only one because looser criteria on
\dedx and \pt can be requested in the former.

For both offline selections, candidates were preselected by requiring
\pt\ (as measured in the inner tracker) to be
greater than $45$\GeVc, the relative
uncertainty in the $\pt$ to be smaller than 0.25, the track fit
$\chi^2/\mathrm{ndf} < 5$, $|\eta| < 1.5$, and the impact parameter
$\sqrt{d_z^2 +d_{xy}^2} < 0.5$\unit{cm}, where $d_z$ and $d_{xy}$ are the
longitudinal and transverse
impact parameters with respect to the reconstructed primary vertex
that yields the smallest track $d_z$ value.
The $\eta$ requirement results from a search optimization
based on the best discovery reach, described in section~\ref{sec:background}.
Candidates were required to have at least two measurements in the silicon pixel
detectors and at least eleven measurements in the inner tracking
detectors before the cleaning procedure.
No more than 20\% of the inner tracker layers were allowed to be
missing between the first and last measurements of the track.
Candidates were required to have at least
six silicon strip measurements passing the cleaning procedure criteria and,
therefore, used for the \dedx and mass measurements.
Candidate tracks were required to have $\ih > 3\MeV\unit{cm}^{-1}$ for the
initial selection.
For the tracker+TOF candidates, the additional requirements of
$\beta^{-1} > 1$, where $\beta$ was computed from the TOF, and
$\sigma_{\beta^{-1}} < 0.07$
were applied.  The number of independent measurements used
for the TOF computation was required to be greater than seven.
Track candidates were required to be loosely isolated as measured
by both the inner tracker and the calorimeters.  Inner tracker isolation was
established by considering all tracks whose direction had a distance from the
candidate track direction, $\DR \equiv \sqrt{(\Delta
\varphi)^2+(\Delta \eta)^2} < 0.3$.
The scalar sum of the $\pt$ of these tracks, with the exception of
the candidate track, was required to be less than
50 (100)\GeVc for the tracker-only (tracker+TOF)
selection. Calorimeter isolation was defined as the ratio between the sum
of the energies measured in each ECAL and HCAL tower within a distance
$\DR < 0.3$ from the candidate direction, and the candidate momentum.
This ratio was required to be less than 0.3 (0.6) for the
tracker-only (tracker+TOF) selection.

\begin{figure*}[htbp]
\begin{center}
\includegraphics[width=0.44\linewidth]{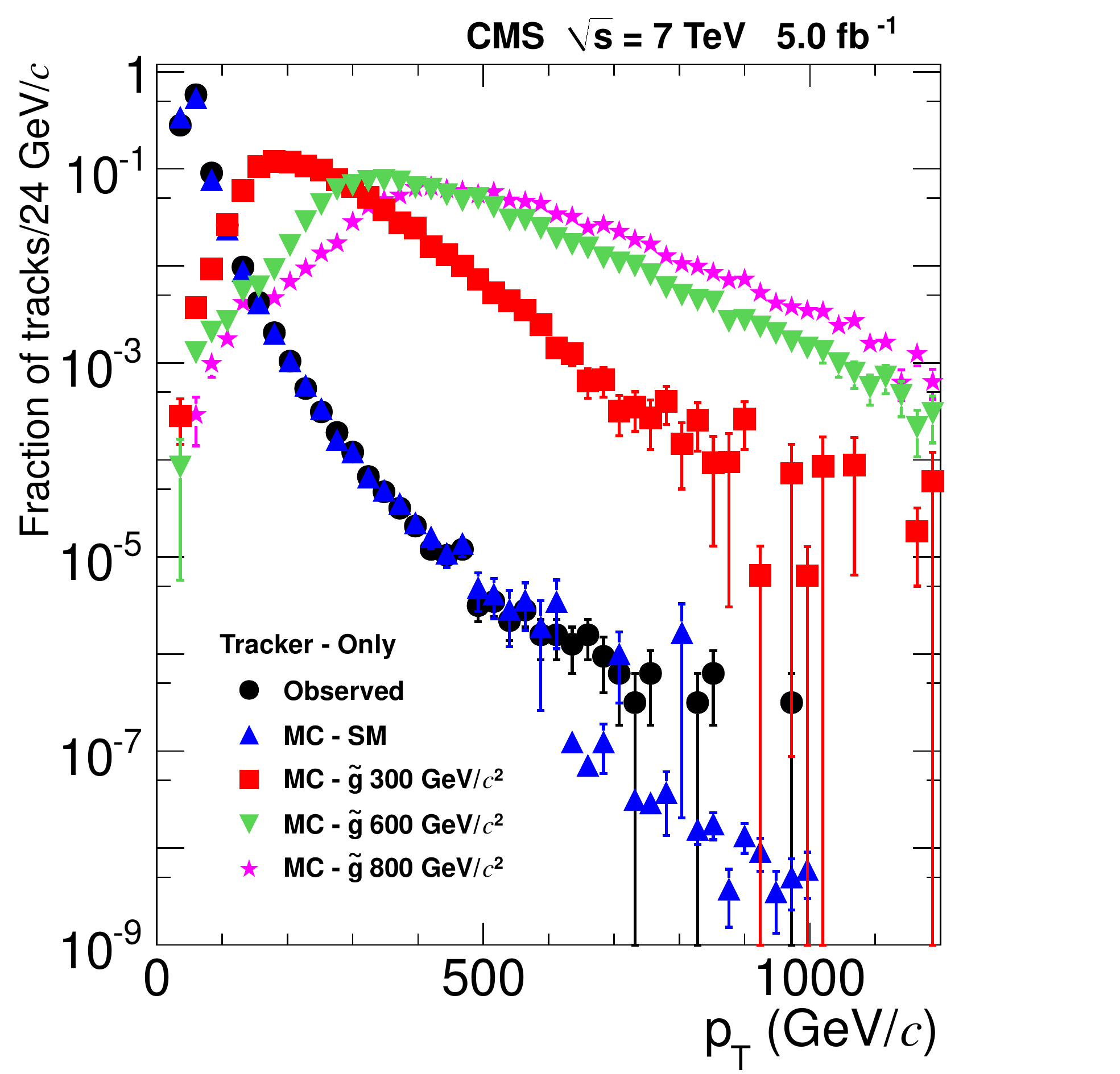}
\includegraphics[width=0.44\linewidth]{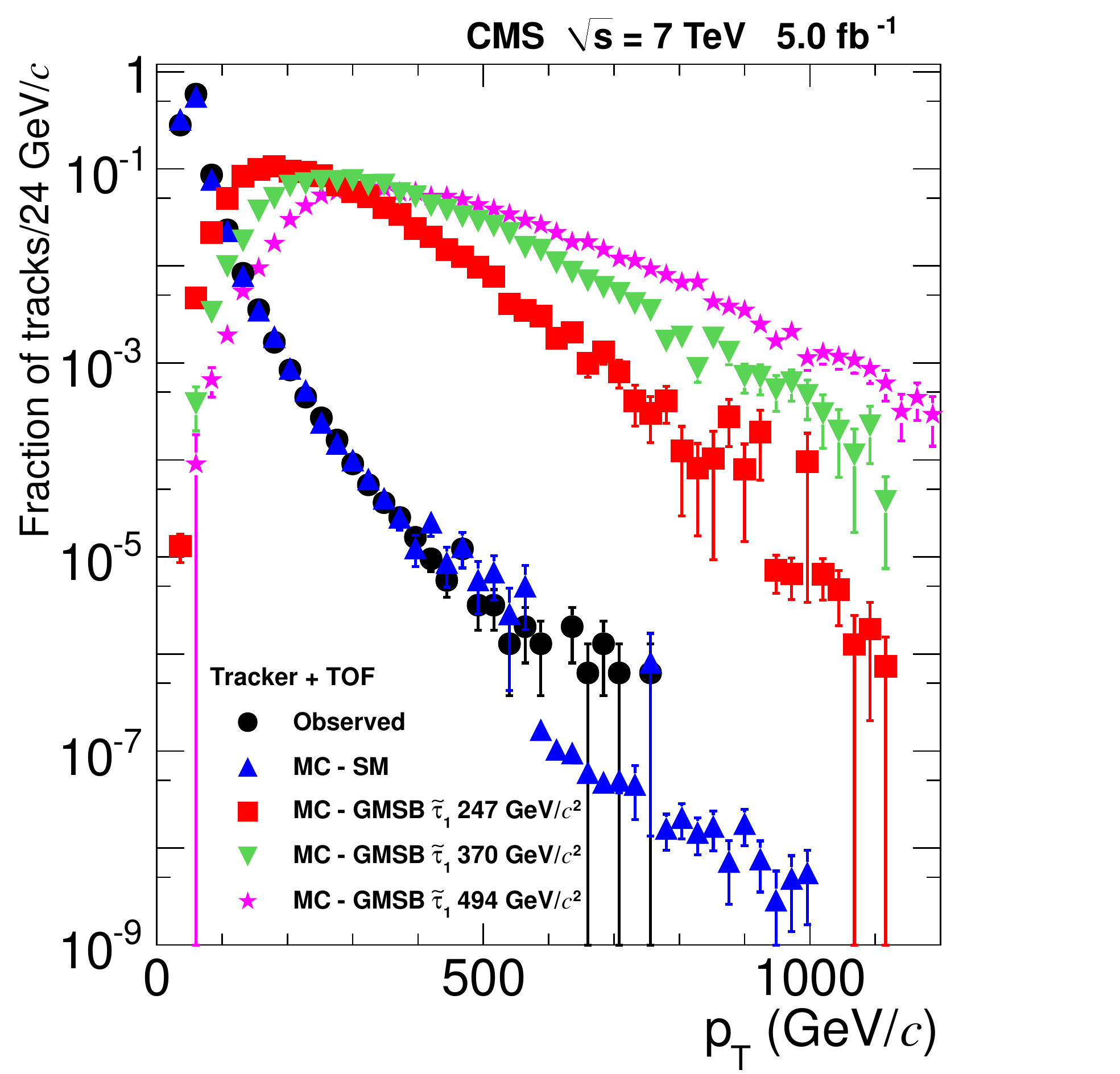}
\includegraphics[width=0.44\linewidth]{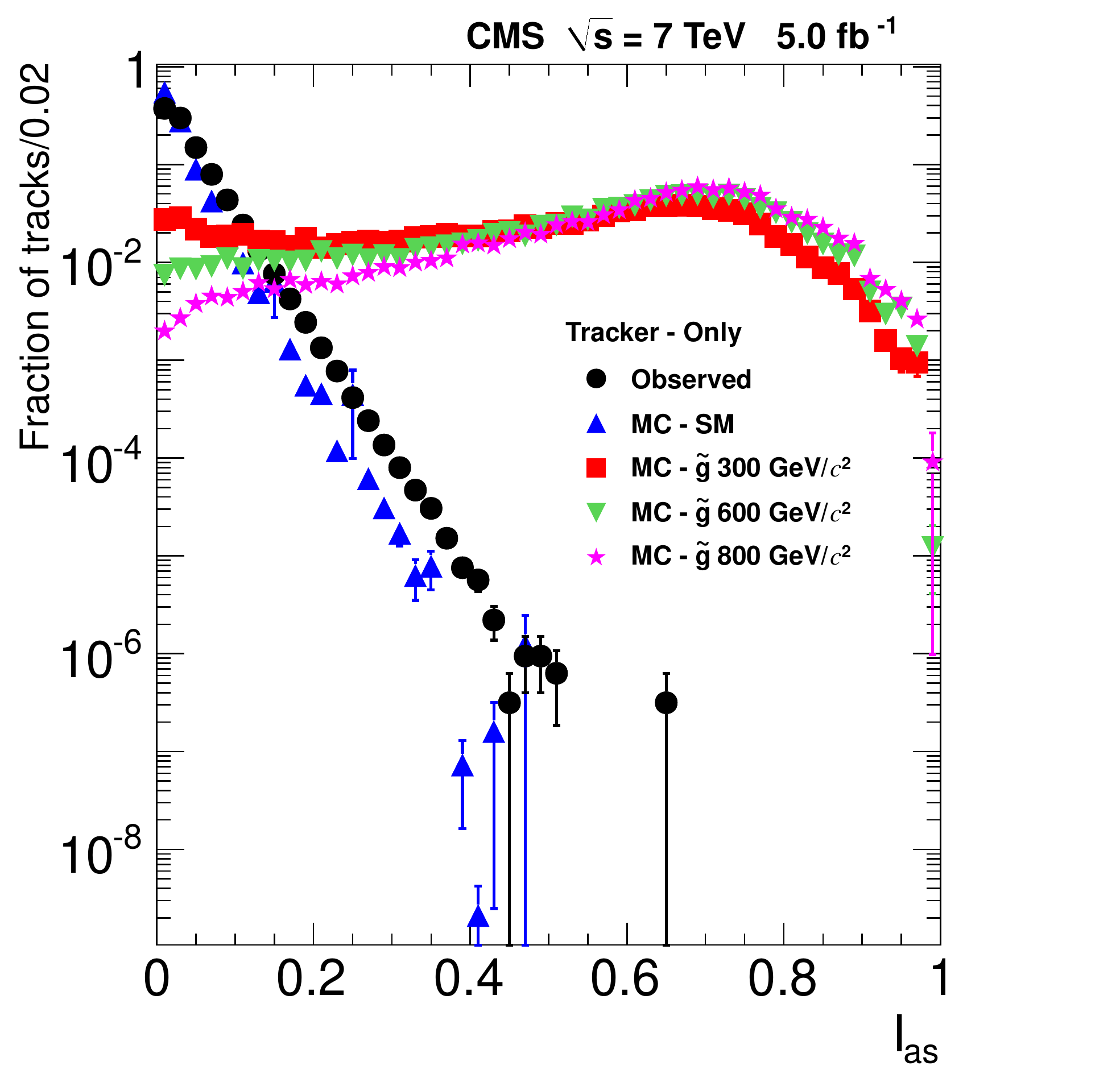}
\includegraphics[width=0.44\linewidth]{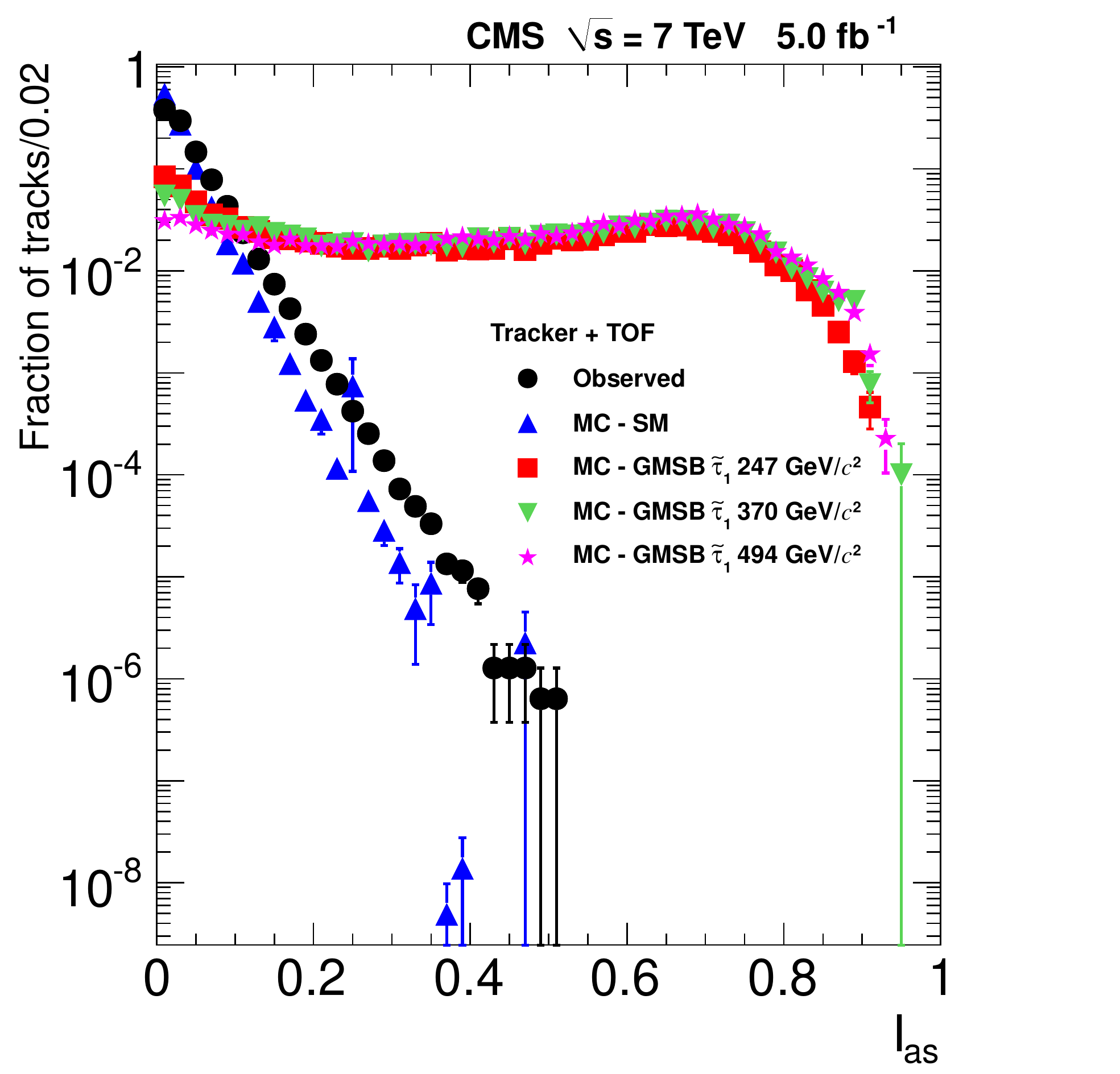}
\hbox to 0.44\linewidth{}\includegraphics[width=0.44\linewidth]{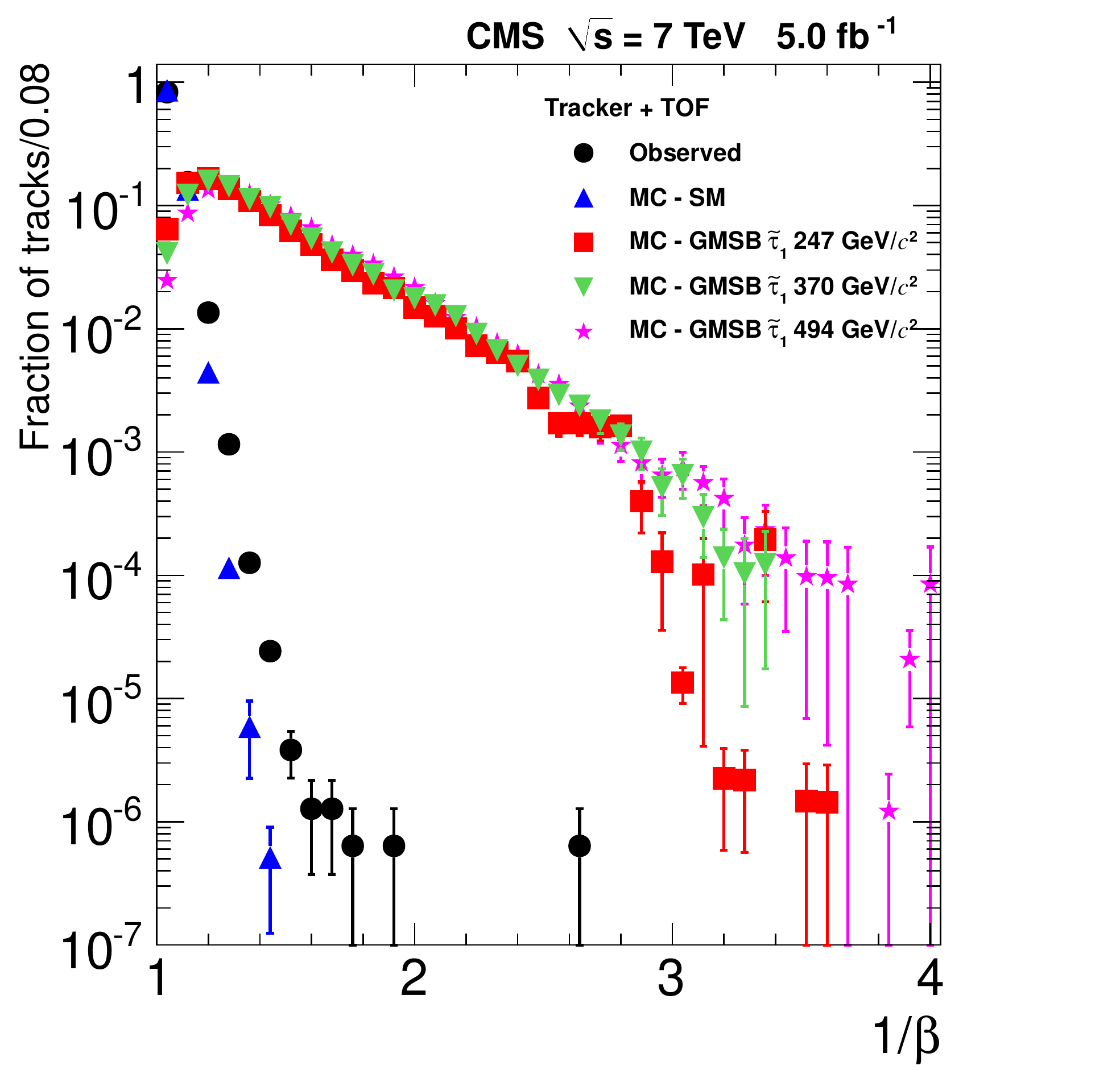}
\caption{Normalized distributions of \pt, \ias, and $\beta^{-1}$ in data,
simulated SM processes, and some of the simulated
signal samples. The two plots on the left are for the tracker-only
selection. The three plots on the right are for the tracker+TOF
selection. Different simulated signal samples are used for the left
and right plots.}
\label{fig:datamc}
\end{center}
\end{figure*}

Good separation between HSCPs and SM particles may be achieved by
selecting candidates with high \pt, high \dedx, and
long TOF (in the tracker+TOF selection). These
quantities are expected to be uncorrelated for SM particles,
while a slow-moving HSCP would have high \dedx and long
TOF even at high \pt. Figure~\ref{fig:datamc} shows
the strong discrimination possible between simulated signals and
background using \ias, TOF, and
\pt. Because of the limited number of available simulated QCD
multi-jet events with low transverse-momentum transfers, which
contribute to the MC distributions for SM processes, these
distributions display bin-to-bin variations in the size of the statistical errors.
A disagreement was found in the tails of the  \ias\ and
$\beta^{-1}$ distributions between the data and the simulation. The
\ias\ discrepancy is understood as due to an increase with time of the
average signal charge observed in the
silicon strip detectors during the 2011 running period. These discrepancies in
the tails have no impact on the estimated background rate since the latter
is determined from data, as described in the
following section. Signal acceptance is instead estimated from MC and
studies were performed to assess the systematic uncertainty
arising from the accuracy of the simulation model of \ias\ and
TOF. They are detailed in section~\ref{sec:results}.

\section{Background determination and search optimization
         \label{sec:background}}

The search was performed as a counting experiment in a mass window that
depended on the HSCP mass hypothesis, $M$, and the model of interest.
For a given $M$, the mass window extended from
$M_\text{reco}-2 \sigma$ to 2\TeVcc, where $M_\text{reco}$ is the average
reconstructed mass for an HSCP of mass $M$ and $\sigma$ is the mass
resolution expected at the true HSCP mass $M$. The values of
$M_\text{reco}$ and $\sigma$ as a function of $M$ were obtained from simulation.

The candidates passing the pre-selection described in
Section~\ref{sec:preselection} were used for both the signal search and
background estimate.
For the tracker-only selection, signal candidates were required to have
\ias\ and \pt\ greater than threshold values optimized for each
model and mass point, as described at the end of this section.
A method that exploits the non-correlation between the
\pt\ and \dedx measurements for SM particles was used to estimate
the background. The number of candidates
expected to pass both the final \pt\ and \ias\ thresholds was
estimated as $D=BC/A$, where $A$ is the number of candidates
that fail both the \ias\ and \pt\ selections and $B$
($C$) is the number of candidates that pass only the \ias\ ($\pt$)
selection.
The $B$ ($C$) candidates were then used to form a binned
probability density function in \ih\ ($p$) for the $D$ candidates such
that, using the mass determination
(Eq.~(\ref{eq:MassFromHarmonicEstimator})), the full
mass spectrum of the background in the signal region $D$ could be
predicted. It was observed that the $\eta$ distribution of the
candidates at low \dedx differs from the distribution
of the candidates at high \dedx. This effect is due to the
typical number of measurements attached to a track, which is
$\eta$-dependent and is anticorrelated with both \ias\ and \ih.
The $\eta$ dependence of \dedx can bias the shape of the predicted
background mass spectrum in the signal region
because the $p$ distribution is also $\eta$-dependent. To
correct for this effect, events in the $C$ region were weighted such that
their $\eta$ distribution matched that in the $B$ region.

For the tracker+TOF selection, this method was
extended to include the TOF
measurement, assuming a lack of correlation between the TOF, $\pt$,
and \dedx measurements.
With three independent and uncorrelated
variables, the number of background candidates in the signal
region may be estimated using six independent combinations
of three out of the eight exclusive samples, each characterized by
candidates passing or failing the three thresholds. These eight samples are
analogous to the $A$, $B$, $C$, and $D$ samples defined above. An
additional independent background estimation in
the signal region $D$ may be obtained with a combination
of four out of the eight samples. The corresponding expression is
$D=AGF/E^2$, where $E$ is the number of candidates
that fail all selections, and $A$, $G$ and $F$
are the numbers of candidates that pass only the \ias, $\pt$, and TOF
selection, respectively.  This latter estimation
has the smallest statistical uncertainty since
the four samples are such that at most one of the three thresholds is
passed in each of them, while the other estimations have at least one
suppressed population sample because of the requirement for two thresholds
to be exceeded.
For this reason the background estimation was
taken from this combination.  As in the tracker-only analysis,
weights were applied to correct the $\eta$ distribution in the regions
providing the \dedx and TOF binned probability density functions that were used to model
the background from SM particles in the signal region.
The dependence of the TOF measurement on $\eta$ for genuine
relativistic muons is due to differences in the typical number
of track measurements, the accuracy of the measurements, the incident angles
of particles on the detectors, and the residual magnetic field in the
muon chamber drift volumes.
The systematic uncertainty in the expected background in the signal
region is estimated to be 10\%, from the differences observed between the
four background estimates having the smallest statistical
uncertainties. The same uncertainty was
adopted
for the tracker-only selection.
The statistical uncertainty of the background estimation in either the
whole signal region or in a given mass range was obtained by generating
simulated pseudoexperiments drawn from the observed distributions in the
control regions.

\begin{figure*}[htbp]
\begin{center}
\includegraphics[width=0.48\textwidth]{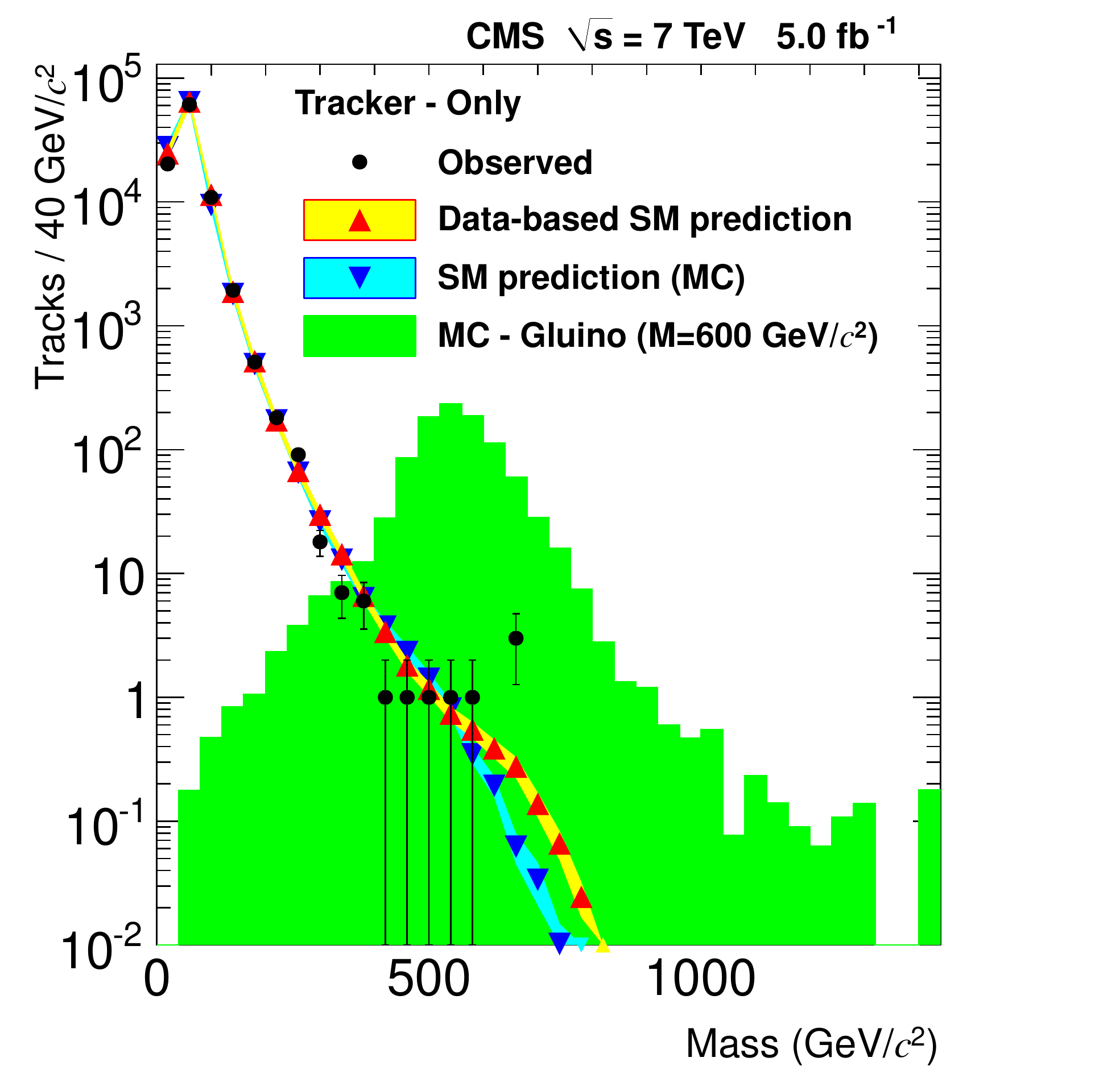}
\includegraphics[width=0.48\textwidth]{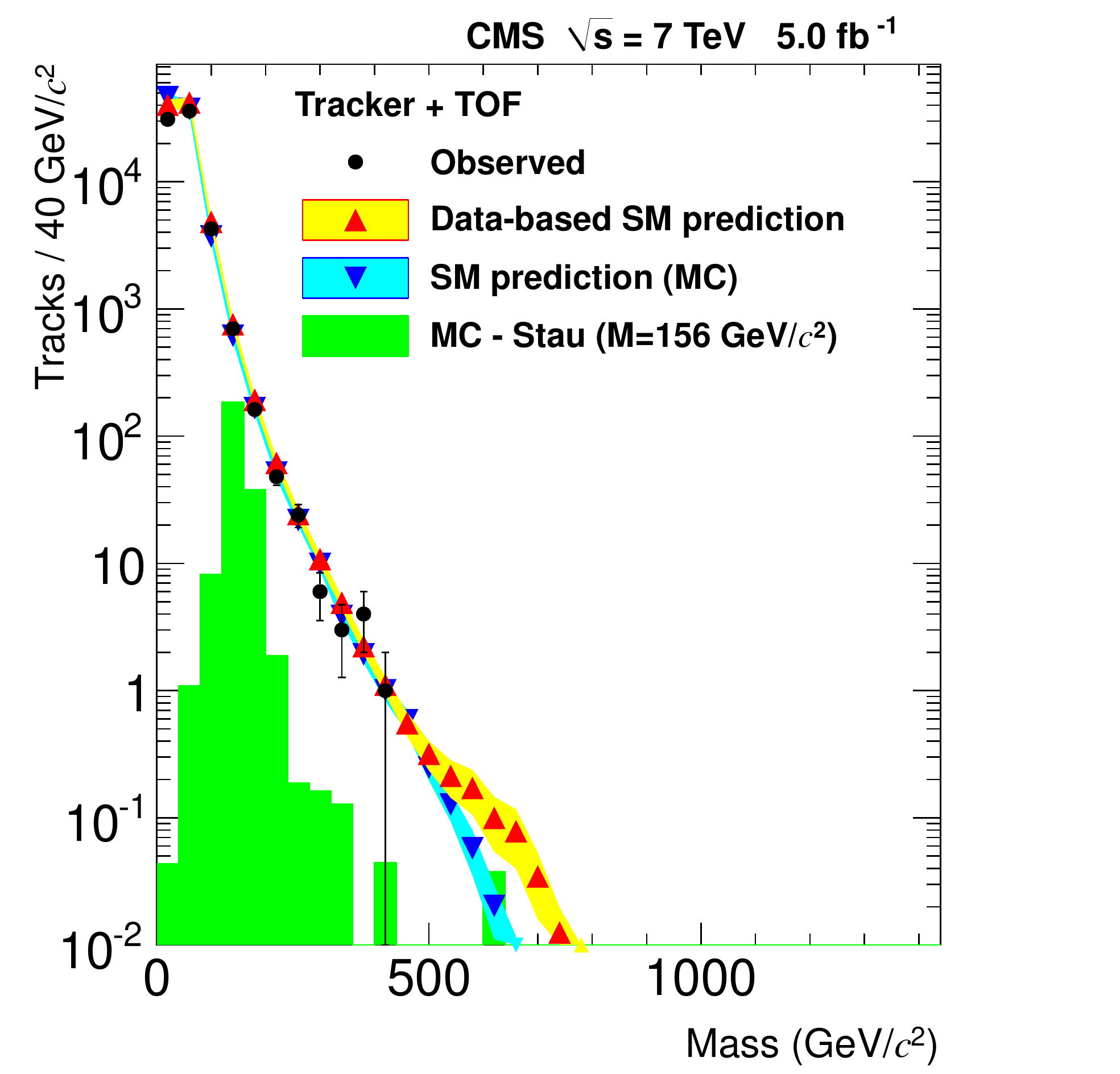}
\caption{Distribution of the candidate mass for the loose selection
defined in Table~\ref{tab:looseselection},
for the tracker-only (left) and tracker+TOF (right) candidates.
Shown are: data, background estimate from data with its uncertainty
(yellow band), simulated signal (green shaded histogram)
and background prediction from MC (blue band) using the same method as
for data. }
\label{fig:ts_bgdshapeprediction}
\end{center}
\end{figure*}

A `loose' selection is defined such that there are a
relatively large number of background candidates in the signal region.
This selection allows a cross-check on the accuracy of the background
prediction to be performed.
Table~\ref{tab:looseselection} reports the minimum values of
\pt, \ias, and $\beta^{-1}$ that candidates must have to pass this
selection, as well as the absolute yields of the background
prediction and the observed data.
Figure~\ref{fig:ts_bgdshapeprediction} shows agreement between
the observed and predicted mass spectra obtained using the loose
selection for both the tracker-only and tracker+TOF candidates. The
background prediction obtained from simulation using the same method as for
data is shown in the same figure.
\begin{table*}[htb]
\begin{center}
\topcaption{\label{tab:looseselection} Selections used to create
the `loose' samples with large number of events and the expected
(Exp.) and observed (Obs.) event yields. The selections are defined in
terms of thresholds in \pt, \ias, and $\beta^{-1}$ (measured from TOF).}
\begin{tabular}{| l | c | c | c | c | c | c | } \hline
Selection     & $\pt^\text{min}$ (\GeVcns) & $I_\mathrm{as}^\text{min}$ & $\beta^{-1}{}^\text{min}$ & Exp.   & Obs.    \\ \hline
Tk-Only   & 50          & 0.10            &  -        &  103450
$\pm$ 10350  (syst) $\pm$ 210 (stat) &  94910     \\ \hline
Tk+TOF  & 50          & 0.05            &      1.05        & 88010
$\pm$ 8800  (syst) $\pm$ 290 (stat) & 72079    \\ \hline
 \end{tabular}
\end{center}
 \end{table*}

The final selection thresholds on \pt, \ias, and TOF were optimized
for each signal model and mass by minimizing the signal cross section
value for which a discovery would be achieved, where discovery is
defined as the expected
mean significance of the observed excess being equal to five standard
deviations with at least five observed candidates.
It was verified that in all cases the optimized
thresholds also guarantee that the expected 95\% confidence level (CL)
cross section upper limit on the considered model is at
most 10\% larger than the minimum attainable.
The optimized selection thresholds and the resulting signal acceptance
for some representative signal models are reported in
Tables~\ref{tab::TABLETKONLY} and~\ref{tab::TABLETKTOF}.

\begin{table*}
\begin{center}
\topcaption{\label{tab::TABLETKONLY} Results of the tracker-only analysis for some
representative signal mass values (in \GeVcc): final selections in terms of minimum values of
\pt (in \GeVc), \ias, $\beta^{-1}$, and $M_\text{reco}$ (in \GeVcc), signal acceptance (``Acc''),
number of candidates expected from SM background (``Exp.''), number of
observed candidates (``Obs.''), predicted theoretical cross
section (``Th. $\sigma$''), expected median cross section upper limit
at 95\% CL for the background-only hypothesis (``Exp. $\sigma$''), and
observed 95\% CL cross section upper limit (``Obs. $\sigma$''). All
cross section values are expressed in pb.}
{\scriptsize
\begin{tabular}{| lr | c c c | c | c  c | c c c |} \hline
 Model                & Mass
&$p_\mathrm{T}^\text{min}$&$I_\mathrm{as}^\text{min}$&$M_\text{reco}^\text{min}$&
Acc.&Exp.&Obs. & Th. $\sigma$& Exp. $\sigma$&Obs. $\sigma$ \\ \hline 
 \PSg\ ($f=0.1$)      &   300 & 60 &   0.400 & 180 &   0.16 &0.328 $\pm$ 0.040 &   0 &   6.6E+01 &  3.8E-03 &   3.7E-03\\ 
 \PSg\ ($f=0.1$)      &  700 &     50 & 0.300 &   410 & 0.21 &  0.089 $\pm$  0.009 &  1 & 2.1E-01 & 3.0E-03 & 4.0E-03 \\  
 \PSg\ ($f=0.1$)      & 1100 &    120 & 0.225  &  570 & 0.15 & 0.094 $\pm$  0.010 &  0 &  3.9E-03 & 4.0E-03 & 3.9E-03 \\ 
 \PSg\ ($f=0.5$)      &  300 &     60 & 0.400 &  180 & 0.086 & 0.328 $\pm$  0.040 &  0 &  6.6E+01 & 6.9E-03 & 6.8E-03\\ 
 \PSg\ ($f=0.5$)      &  700 &     50 & 0.300  &  410 & 0.12 & 0.089 $\pm$  0.009 &  1 &  2.1E-01 & 5.3E-03 & 7.1E-03\\ 
 \PSg\ ($f=0.5$)      & 1100 &    120 & 0.225  &  570 & 0.085 & 0.094 $\pm$  0.010 &  0 &  3.9E-03 & 7.0E-03 & 6.9E-03 \\ 
 \PSg\ ($f=0.1$, ch. suppr.)     &  300 &     60 & 0.400 & 180 & 0.020 & 0.328 $\pm$  0.040 &  0 &  6.6E+01 & 3.0E-02 & 3.0E-02 \\ 
 \PSg\ ($f=0.1$, ch. suppr.)     &  700 &     50 & 0.325 &  370 & 0.045 & 0.092 $\pm$  0.010 &  1 &  2.1E-01 & 1.4E-02 & 1.8E-02 \\ 
 \PSg\ ($f=0.1$, ch. suppr.)     & 1100 &     50 & 0.275  &  460 & 0.032 & 0.085 $\pm$  0.009 &  1 &  3.9E-03 & 1.9E-02 & 2.6E-02 \\ 
  \stone\                 &  200 &     60 & 0.400 &  130 & 0.14 & 1.250 $\pm$  0.160 &  4 &  1.3E+01 & 5.8E-03 & 1.1E-02 \\
  \stone\                 &  500 &     50 & 0.350  &  310 & 0.24 & 0.126 $\pm$  0.014 &  0 &  4.8E-02 & 2.4E-03 & 2.3E-03 \\ 
  \stone\                 &  800 &     50 & 0.275  &  450 & 0.29 & 0.095 $\pm$  0.010 &  1 &  1.1E-03 & 2.1E-03 & 2.8E-03 \\ 
  \stone\ (ch. suppr.)    &  200 &     70 & 0.400 & 120 & 0.021 & 1.520 $\pm$0.202 &  4 &  1.3E+01 & 4.0E-02 & 7.2E-02 \\
  \stone\ (ch. suppr.)    &  500 &     50 & 0.375  &  280 & 0.064 &  0.102 $\pm$  0.012 &  0 & 4.8E-02 & 9.1E-03 & 9.1E-03 \\ 
  \stone\ (ch. suppr.)    &  800 &     50 & 0.325  &  370 & 0.077 & 0.092 $\pm$  0.010 &  1 &  1.1E-03 & 8.1E-03 & 1.1E-02 \\ 
 GMSB \stau\            &  100 &     65 & 0.400 &  20 &  0.12 & 6.980 $\pm$  0.908 &  7 & 1.3E+00 & 1.2E-02 & 1.3E-02 \\ 
 GMSB \stau\            &  494 &     65 & 0.350 &  300 & 0.64 & 0.126 $\pm$  0.014 &  0 &  6.2E-05 & 9.3E-04 & 9.3E-04 \\ 
 pair prod. \stau\      &  100 &     70 & 0.400 &   40 & 0.11 &  4.840 $\pm$  0.608 &  6 & 3.8E-02 & 1.2E-02 & 1.5E-02 \\ 
 pair prod. \stau\      &  308 &     70 & 0.400 &  190 & 0.39 &  0.237 $\pm$  0.030 &  0 & 3.5E-04 & 1.5E-03 & 1.5E-03 \\ 
 \hyperk\    (\hyperrho(800))     &  100 &     70 & 0.400 &   10 &  0.065 & 4.880 $\pm$  0.613 &  6 & 1.4E+00 & 1.9E-02 & 2.3E-02 \\ 
 \hyperk\    (\hyperrho(800))     &  500 &     50 & 0.350  &  320 &  0.61 & 0.107 $\pm$  0.012 &  0 & 2.8E-04 & 9.6E-04 & 9.6E-04 \\ 
 \hyperk\    (\hyperrho(1200))     &  600 &     50 & 0.325 &  370 &  0.22 & 0.092 $\pm$  0.010 &  1 & 2.6E-03 & 2.8E-03 & 3.8E-03 \\ 
 \hyperk\    (\hyperrho(1200))     &  700 &     50 & 0.275  &  440 &  0.65 & 0.106 $\pm$  0.011 &  1 & 6.1E-05 & 9.6E-04 & 1.3E-03 \\ 
 \hyperk\    (\hyperrho(1600))     &  800 &    140 & 0.250 &  480 & 0.33 &  0.118 $\pm$  0.012 &  1 & 2.6E-04 & 1.9E-03 & 2.5E-03 \\ 
 \hyperk\    (\hyperrho(1600))     &  900 &    135 & 0.225 &  530 & 0.62 &  0.128 $\pm$  0.014 &  0 & 1.3E-05 & 9.3E-04 & 9.3E-04 \\ 
\hline
\end{tabular}}
\end{center}
\end{table*}

\begin{table*}
\begin{center}
\topcaption{\label{tab::TABLETKTOF} Results of the tracker+TOF analysis for some
representative signal mass values (in \GeVcc): final selections in terms of minimum values of
\pt (in \GeVc), $I_\mathrm{as}$, $\beta^{-1}$, and $M_\text{reco}$ (in \GeVcc), signal acceptance (``Acc''),
number of candidates expected from SM background (``Exp.''), number of
observed candidates (``Obs.''), predicted theoretical cross
section (``Th. $\sigma$''), expected median cross section upper limit
at 95\% CL for the background-only hypothesis (``Exp. $\sigma$''), and
observed 95\% CL cross section upper limit (``Obs. $\sigma$''). All
cross section values are expressed in pb.}
{\scriptsize
\begin{tabular}{| lr | c c c c | c c | c | c c c |} \hline
 Model                & Mass
&$p_\mathrm{T}^\text{min}$&$I_\mathrm{as}^\text{min}$&$\beta^{-1}{}^\text{min}$&$M_\text{eco}^\text{min}$&Acc.&Exp.&Obs. &
Th. $\sigma$& Exp. $\sigma$&Obs. $\sigma$  \\ \hline
 \PSg\ ($f=0.1$)      &  300 &     55 & 0.175 & 1.175 &  180 & 0.17 &  0.119 $\pm$  0.012 &  0 & 6.6E+01 & 3.4E-03 & 3.4E-03 \\ 
 \PSg\ ($f=0.1$)      &  700 &    110 & 0.050 & 1.125 &  430 & 0.19 &  0.113 $\pm$  0.015 &  0 & 2.1E-01 & 3.0E-03 & 3.0E-03 \\ 
 \PSg\ ($f=0.1$)      & 1100 &    110 & 0.025 & 1.075 &  620 & 0.13 &  0.111 $\pm$  0.033 &  0 & 3.9E-03 & 4.6E-03 & 4.6E-03 \\ 
 \PSg\ ($f=0.5$)      &  300 &     55 & 0.175 & 1.175 &  180 & 0.094 &  0.119 $\pm$  0.012 &  0 & 6.6E+01 & 6.3E-03 & 6.2E-03 \\ 
 \PSg\ ($f=0.5$)      &  700 &    110 & 0.050 & 1.125 &  430 & 0.11 &  0.113 $\pm$  0.015 &  0 & 2.1E-01 & 5.4E-03 & 5.3E-03 \\ 
 \PSg\ ($f=0.5$)      & 1100 &    110 & 0.025 & 1.075 &  620 & 0.072 &  0.111 $\pm$  0.033 &  0 & 3.9E-03 & 8.2E-03 & 8.2E-03 \\ 
\stone\            &  200 &     50 & 0.200 & 1.200 &  130 &  0.15 &0.109$\pm$  0.011 &  0 &  1.3E+01 & 3.9E-03 & 3.8E-03 \\
  \stone\                 &  500 &     60 & 0.075 & 1.150 &  330 & 0.25 &  0.125 $\pm$  0.013 &  0 & 4.8E-02 & 2.4E-03 & 2.4E-03 \\ 
  \stone\                 &  800 &    105 & 0.025 & 1.125 &  490 & 0.26 &  0.096 $\pm$  0.019 &  0 & 1.1E-03 & 2.2E-03 & 2.2E-03 \\ 
 GMSB \stau\            &  100 &     50 & 0.300 & 1.275 &   30 & 0.20 &  0.093 $\pm$  0.011 &  0 & 1.3E+00 & 2.9E-03 & 2.9E-03 \\ 
 GMSB \stau\            &  494 &     55 & 0.025 & 1.175 &  320 & 0.78 &  0.113 $\pm$  0.014 &  1 & 6.2E-05 & 7.8E-04 & 1.1E-03 \\ 
 pair prod. \stau\      &  100 &     50 & 0.250 & 1.275 &   50 & 0.19 &  0.109 $\pm$  0.012 &  0 & 3.8E-02 & 3.0E-03 & 2.9E-03 \\ 
 pair prod. \stau\      &  308 &     65 & 0.125 & 1.200 &  190 & 0.55 &  0.105 $\pm$  0.011 &  0 & 3.5E-04 & 1.1E-03 & 1.1E-03 \\ 
\hyperk\    (\hyperrho(800))     &  100 &     50 & 0.300 & 1.275 &   20 & 0.11 &  0.095 $\pm$  0.011 &  0 & 1.4E+00 & 5.3E-03 & 5.2E-03 \\ 
 \hyperk\    (\hyperrho(800))     &  500 &     60 & 0.075 & 1.150 &  330 & 0.68 &  0.125 $\pm$  0.013 &  0 & 2.8E-04 & 8.6E-04 & 8.5E-04 \\ 
 \hyperk\    (\hyperrho(1200))     &  600 &     70 & 0.025 & 1.150 &  380 & 0.22 &  0.107 $\pm$  0.015 &  0 & 2.6E-03 & 2.6E-03 & 2.6E-03 \\ 
 \hyperk\    (\hyperrho(1200))     &  700 &    110 & 0.050 & 1.125 &  450 & 0.66 &  0.087 $\pm$  0.013 &  0 & 6.1E-05 & 9.0E-04 & 9.0E-04 \\ 
 \hyperk\    (\hyperrho(1600))     &  800 &     50 & 0.050 & 1.100 &  500 & 0.33 &  0.119 $\pm$  0.021 &  0 & 2.6E-04 & 1.8E-03 & 1.8E-03 \\ 
 \hyperk\    (\hyperrho(1600))     &  900 &     85 & 0.075 & 1.075 &  550 & 0.61 &  0.123 $\pm$  0.022 &  0 & 1.3E-05 & 9.3E-04 & 9.1E-04 \\ 
\hline
\end{tabular}}
\end{center}
\end{table*}

\section{Results \label{sec:results}}

After comparing data in the signal region with the expected
background for all optimized selections, no
statistically significant excess was observed.
Tables~\ref{tab::TABLETKONLY} and~\ref{tab::TABLETKTOF} report results for some
representative selections.
The largest excess has a significance of 1.75 one-sided Gaussian
standard deviations and was found with the selection optimized for a
\stone\ with a mass of 200\GeVcc.  Only one of the three highest mass candidates
passing the tracker-only loose selection
(Fig.~\ref{fig:ts_bgdshapeprediction}) passes one of the final
selections. This candidate is also associated with an identified muon
and has $\beta^{-1} = 1.03$, which is well below any threshold used in
the tracker+TOF final selections.

The observed data sample was used to calculate upper limits on the HSCP production cross section for
each considered model and mass point. The cross section upper limits
at 95\% CL were obtained using a CL$_s$ approach~\cite{Read:2000ru}
with a one-sided profile likelihood test statistic whose p-values were evaluated
by generating pseudoexperiments using a frequentist
prescription~\cite{Chatrchyan:2012tx}.
A log-normal probability density function~\cite{James} was used for the
nuisance parameter measurements,
which are the integrated luminosity, the signal acceptance, and the
expected background yield in the signal region. When
generating pseudoexperiments for the limit calculation, each nuisance
parameter was drawn from the corresponding probability density
function with a central value equal to the
best fit value to data under the signal+background hypothesis.
All systematic uncertainties are summarized in
Table~\ref{tab::systematicerror} and are incorporated in the limits
quoted below.

Simulation was used to determine the signal acceptance.
A number of studies were undertaken to estimate the degree to which the
simulation correctly models the detector response to HSCPs and to
assess an uncertainty in the signal acceptance.

The uncertainty in the trigger efficiency derived from simulation
was evaluated separately for the \MET\ and single-muon triggers.
The uncertainty in the \MET\ trigger efficiency was
dominated by the uncertainty in the jet-energy
scale~\cite{Chatrchyan:2011ds}, which was less than 3\% across the
energy range.
For the charge-suppression models, where the \MET\ trigger
is most relevant, varying the jet-energy scale and jet-energy resolution
within their uncertainty resulted in a relative change of the
trigger efficiency by no more than 5\%.
For all the other models, which made use of
overlap with the single-muon trigger, the relative change of the overall
trigger efficiency was found to vary by no more than 2\%.
For the single-muon trigger, a relative disagreement of up to 5\% was
observed between the efficiency estimated in data and MC at
all energies~\cite{MUO-10-002}.
In addition, for this specific analysis, a further
uncertainty arises from the imperfect simulation of the synchronization of
the muon trigger electronics. The accuracy of the synchronization
was evaluated from data separately for each muon subdetector.
This effect was found to yield less than 2\% relative uncertainty
on the overall trigger efficiency for all considered signals.
On the basis of these numbers, an uncertainty of 5\% on
the overall trigger efficiency was assumed for all models.

The accuracy of the \dedx model used in simulation was studied using
low-momentum protons and kaons. The simulation was
found to underestimate both the
\ih\ and \ias\ scales by less than 5\%. The \ias\ resolution was
in contrast found to be overestimated by a constant value of 0.08 in the
region around the thresholds adopted in the analysis. After corrections
for these discrepancies were applied to simulation,
only 20\% of the signal models
displayed an efficiency decrease, with the maximum relative reduction
being smaller than 2\%. The efficiency for all other models increased by
up to 10\% relative to the uncorrected MC result. Based on these
results, the efficiency determined from the simulation was not corrected,
but was assigned an associated uncertainty of 2\%.

The accuracy of the TOF model implemented in the simulation was studied using
cosmic ray muons and muons produced directly in collisions.
In the region around the  $\beta^{-1}$
thresholds adopted in the analysis, the simulation was found to
overestimate $\beta^{-1}$ by a constant value of 0.003
and 0.02 for DT and CSC, respectively. The resolution of the
measurement of $\beta^{-1}$ was found to be well modelled in
simulation. After corrections for these discrepancies were applied,
the signal efficiency was found to decrease by no more than 2\% for
all considered models and mass points. This maximum change of 2\% was
adopted as the uncertainty associated with the TOF measurement.

The uncertainty on the track momentum scale was modelled by
varying the track \pt as a function of the track $\phi$ and $\eta$
values such that~\cite{MUO-10-004}:
\begin{equation}
\frac{1}{\pt^\prime} = \frac{1}{\pt} + \delta_{K_T}(q, \phi, \eta) ,
\label{eq:shift1}
\end{equation}

\begin{equation}
\delta_{K_T}(q, \phi, \eta) = a + b\eta^2 + qd\sin(\phi - \phi_0) ,
\label{eq:shift2}
\end{equation}

where $q$ is the track charge sign ($q=\pm 1$) and the function
$\delta_{K_T}$ controls the shift in the track momentum scale. This
function has four free parameters, $a$, $b$, $d$, and $\phi_0$, whose
values were obtained~\cite{MUO-10-004} by minimizing the difference
between the invariant mass distributions of $Z \rightarrow \mu^+
\mu^-$ candidates in data and simulation. The obtained values are $a =
0.236\TeV^{-1}c$, $b = -0.135\TeV^{-1}c$, $d = 0.282\TeV^{-1}c$,
and $\phi_0 = 1.337$ rad. The phi dependence is believed to be due to
imperfect inner tracker alignment. The expected shift in
inverse \pt for tracks of higher momenta measured in the inner tracker
are found~\cite{MUO-10-004} to be compatible with those provided by
equations~\ref{eq:shift1} and~\ref{eq:shift2}.
The difference between the signal acceptance with the nominal and shifted
\pt\ was taken as the uncertainty and was found to be smaller than 4\%.

The uncertainties in the efficiencies for reconstructing
muons~\cite{MUO-10-002}, and for reconstructing tracks in the inner
tracker~\cite{TRK-10-002}  were also considered and established to be
less than 2\% in each case.

The impact of the uncertainty in the mean rate of additional
interactions in each bunch crossing
was studied and found to be negligible compared to the statistical precision
(0.5\%) allowed by the size of the simulated signal samples.

Two theoretical uncertainties affecting the signal acceptance
were studied: the uncertainty in the model of hadronization
and nuclear interactions, and the uncertainty due to the
MPI tune.  The hadronization and nuclear-interaction model
is discussed in Section~\ref{sec:signalmc}.
Results are obtained
for two very different nuclear interaction models and for two different
\PSg\ hadronization schemes. With regard to the MPI tune,
tune Z2 uses a \pt-ordered model, which
appears to generate significantly more initial-state radiation than the
$Q^2$-ordered D6T model. For some models a significant increase in
the trigger efficiency and in the reconstruction efficiency is found
and the
observed limits become more stringent.
The most conservative set of limits, resulting from the $Q^2$-ordered
D6T model, are those reported.

An uncertainty of 2.2\% is estimated~\cite{SMP-12-008} for the absolute
value of the integrated luminosity.
The uncertainty in the expected background was discussed in
Section~\ref{sec:background} and is estimated to be of the order of 10\%.
This uncertainty has very little impact on the results, because of the
small numbers of expected events for most mass points.

\begin{table}
\begin{center}
\topcaption{\label{tab::systematicerror} Sources of systematic
uncertainties and corresponding relative uncertainties.}
\begin{tabular}{| l | c | } \hline
Source of systematic uncert. & Relative uncert. ($\%$)  \\
\hline \hline
Signal acceptance: &  \\
~-~Trigger efficiency & 5  \\
~-~Track momentum scale & $< 4$  \\
~-~Ionization energy loss & 2  \\
~-~Time-of-flight & 2  \\
~-~Track reconstruction eff. & $<2$  \\
~-~Muon reconstruction eff. & $<2$  \\
~-~Pile-up & $< 0.5$  \\
Total uncert. in signal acc. & 7  \\ \hline

Expected background  &  10  \\ \hline
Integrated luminosity & 2.2  \\ \hline
 \end{tabular}
\end{center}
\end{table}

\begin{figure}[htbp]
\begin{center}
\includegraphics[width=0.48\textwidth]{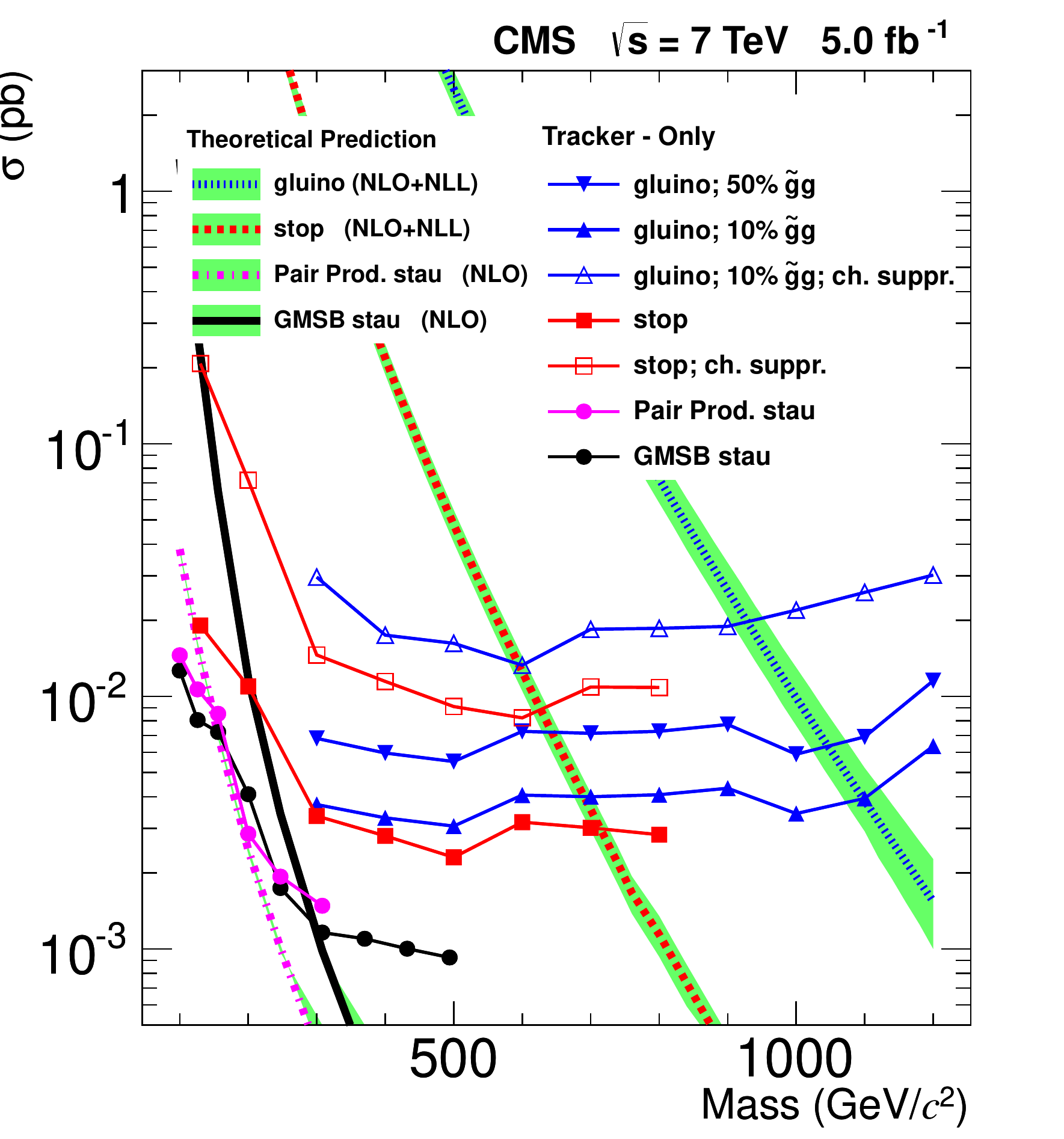}
\includegraphics[width=0.48\textwidth]{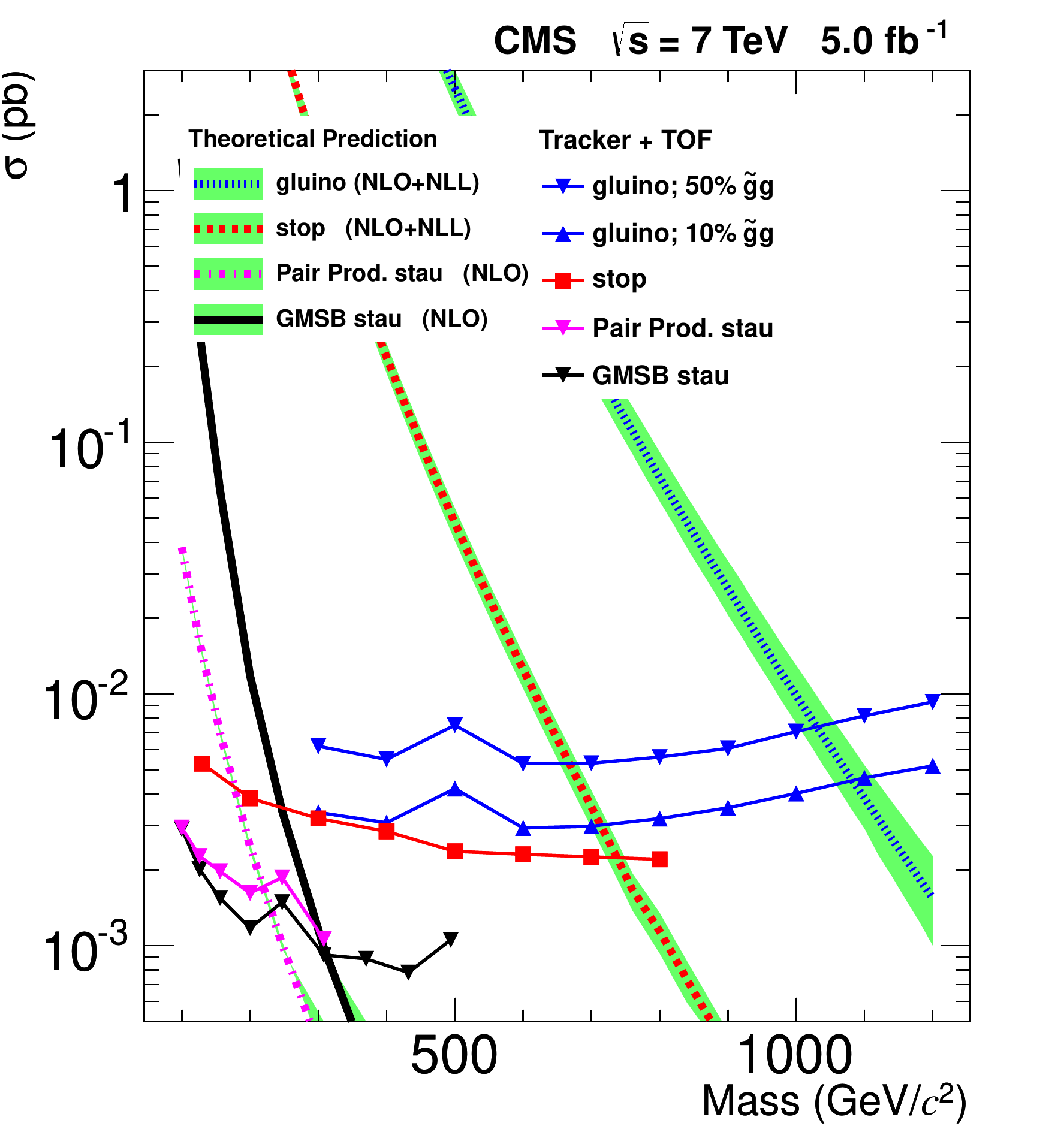}
\caption{\label{fig::STOPMassExclusion} Predicted theoretical cross
section and observed 95\% CL upper limits on the cross section for
the different signal models considered:
production of \stone, \PSg, and \stau;
different fractions $f$ of R-gluonball states produced after hadronization;
standard and charge suppression (ch. suppr.) scenario. \cmsLeft:
tracker-only selection. \cmsRight: tracker+TOF. The uncertainties in
the theoretical cross section are shown as bands around
the central values.}
\end{center}
\end{figure}

\begin{figure}[htbp]
\begin{center}
\includegraphics[width=0.48\textwidth]{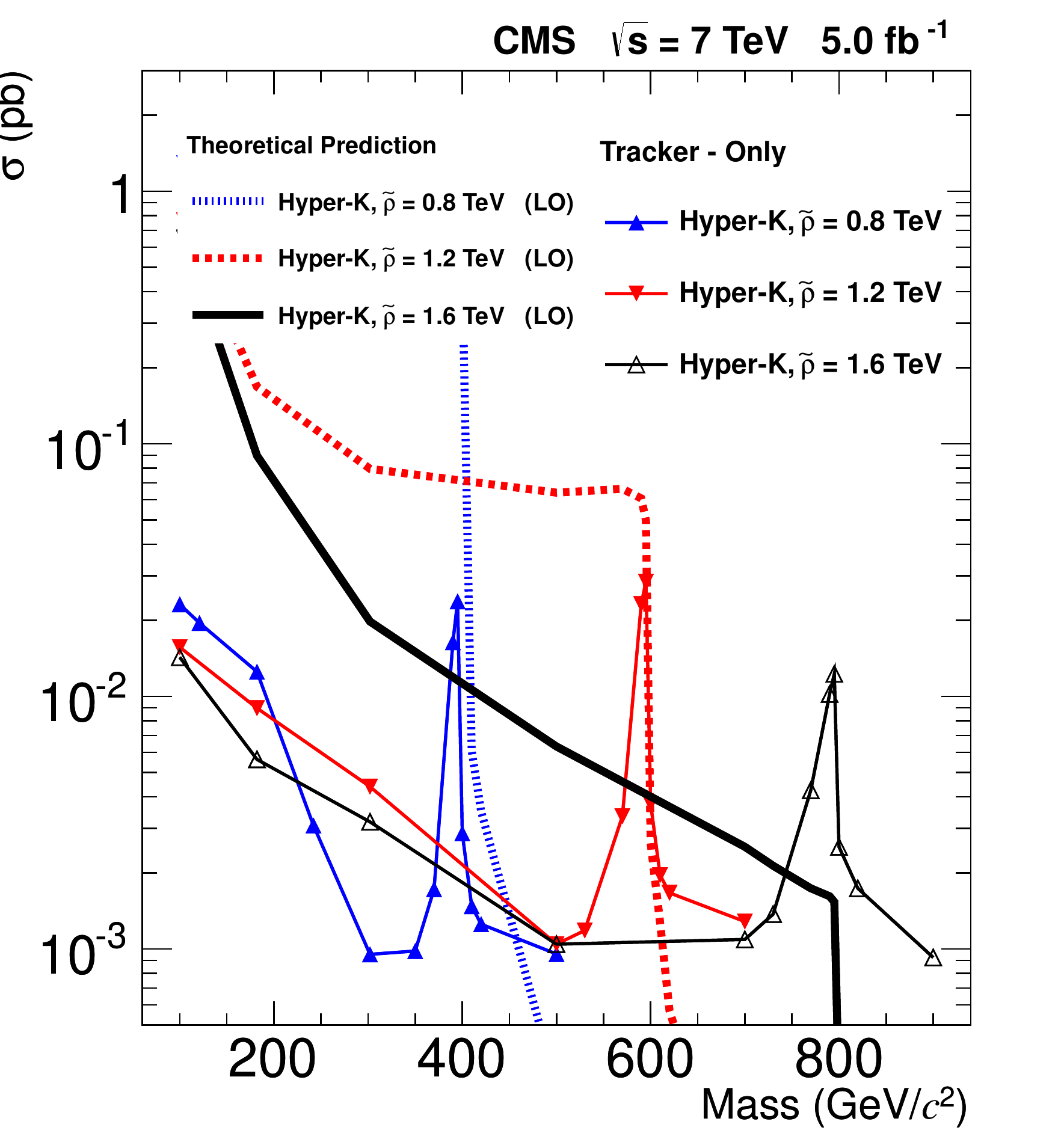}
\includegraphics[width=0.48\textwidth]{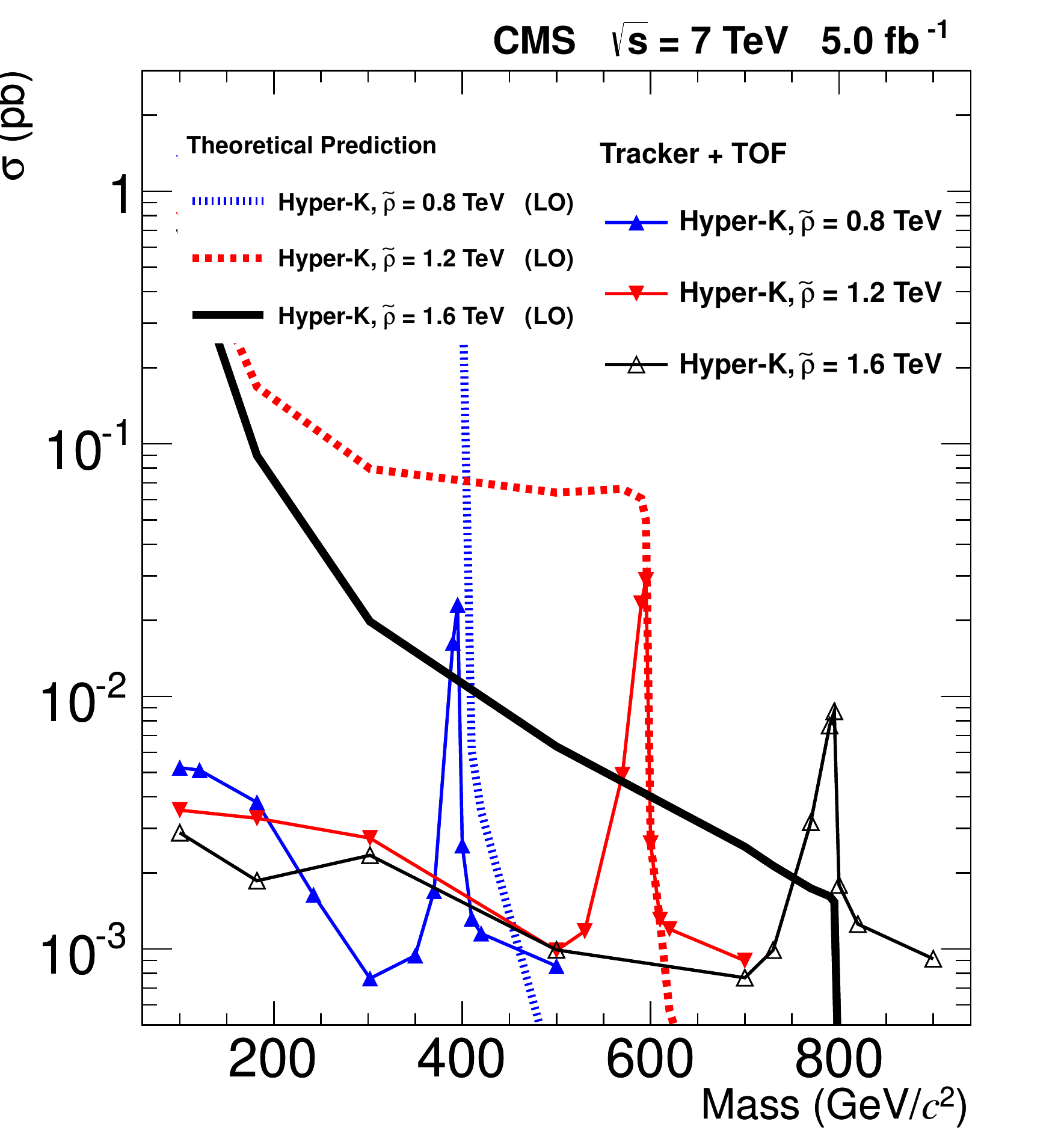}
\caption{\label{fig::DiChampMassExclusion} Predicted theoretical cross
section and observed 95\% CL upper limits on the cross section for
the \hyperk\ models with different \hyperrho\ mass values. \cmsLeft:
tracker-only selection. \cmsRight: tracker+TOF.  }
\end{center}
\end{figure}

\begin{figure*}[htbp]
\begin{center}
\includegraphics[width=0.48\textwidth]{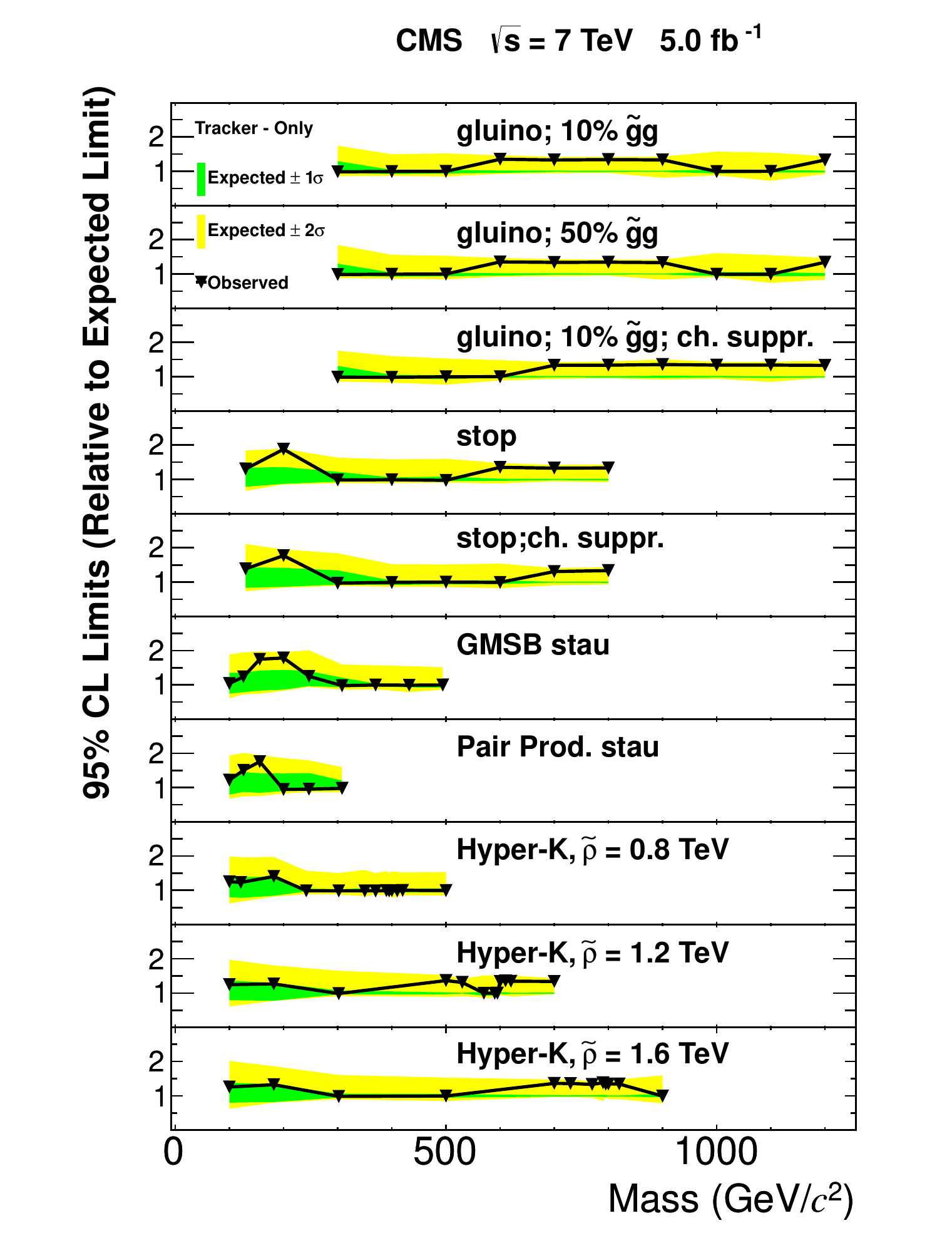}
\includegraphics[width=0.48\textwidth]{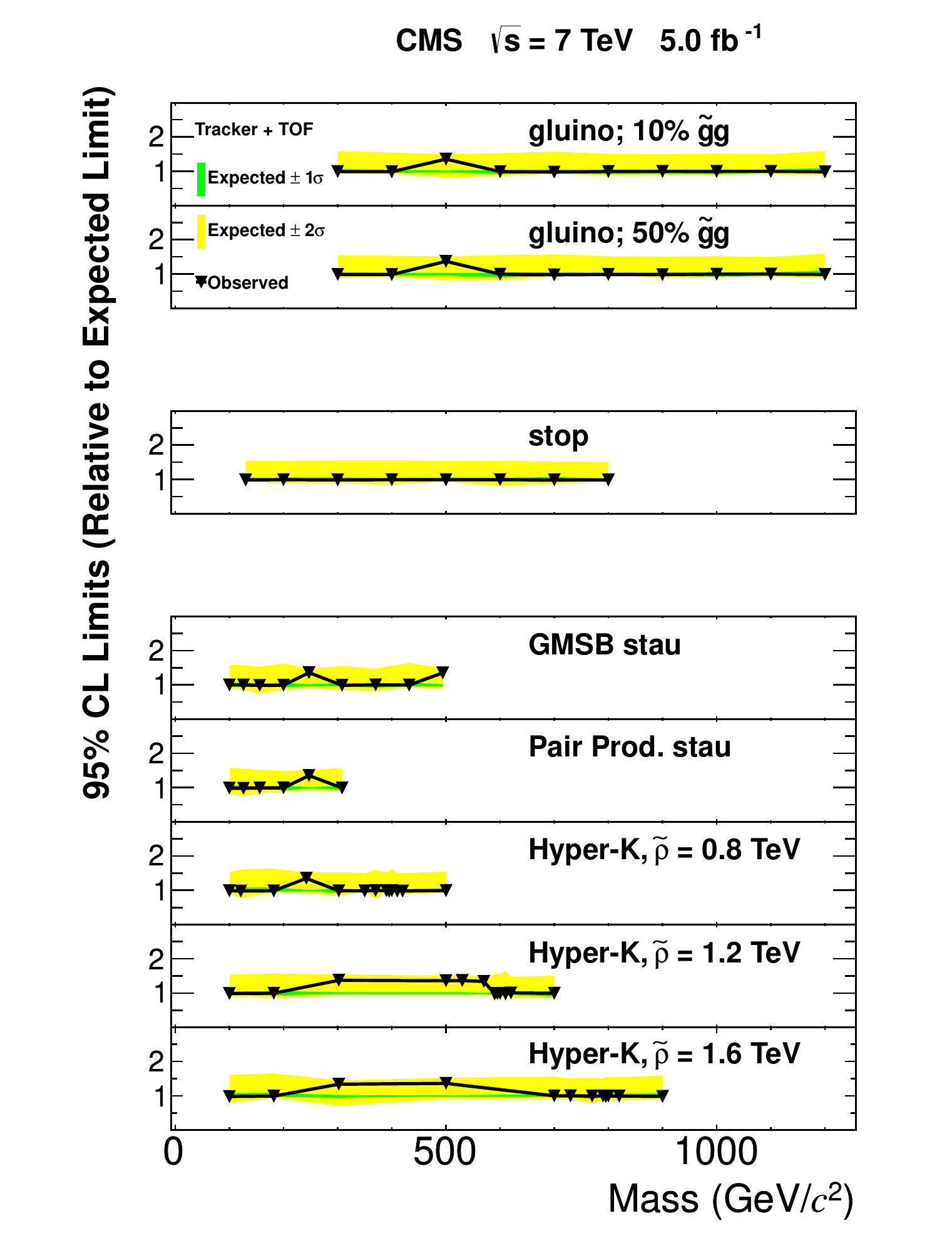}
\caption{\label{fig::LimitsRatio} Ratio of observed
95\% CL upper limits to expected median limits for the background-only
hypothesis. The green (dark) and yellow
(light) bands indicate the ranges that are expected to contain
68\% and 95\% of all observed excursions from the expected median,
respectively. Ratios are presented for the different signal models considered:
production of \PSg, \stone, \stau, and \hyperk;
different fractions, $f$, of R-gluonball states produced after hadronization;
standard or charge suppression (ch. suppr.) scenario. Left: tracker-only
selection. Right: tracker+TOF.}
\end{center}
\end{figure*}

The 95\% CL cross section upper limit curves obtained with both the
tracker-only and the tracker+TOF selection
are shown in Figs.~\ref{fig::STOPMassExclusion}
and~\ref{fig::DiChampMassExclusion},
along with the theoretical predictions for the chosen signal
models.
The ratio of observed to expected 95\% CL upper limits on the cross
section is reported in Fig.~\ref{fig::LimitsRatio}  for
the different combinations of models and scenarios considered.
Numerical values for the predicted theoretical cross
section, and expected and observed cross section upper limit
at 95\% CL are reported in Tables~\ref{tab::TABLETKONLY}
and~\ref{tab::TABLETKTOF} for some representative signal models.
For \stone\ and \PSg\ pair
production, theoretical cross sections were computed at
next-to-leading order (NLO) plus next-to-leading-logarithmic (NLL)
accuracy~\cite{Kulesza:2008jb,Kulesza:2009kq, Beenakker:2009ha,
Beenakker:2010nq,Beenakker:2011fu} using \PROSPINO
v2.1~\cite{Beenakker:1996ed}.
The uncertainty in these theoretical cross section values
varies between 10\% to 25\% and is shown in
Fig.~\ref{fig::STOPMassExclusion} as a band around the central value.
The cross sections for the models with \stau\ production were
calculated at NLO with \PROSPINO v2.1.  The uncertainty in the
theoretical cross section was estimated to be 5\% to 14\%  for
the GMSB model and 3\% to 7\% for direct \stau\ pair production,
depending on the mass.
In all cases the sources of uncertainty include
renormalization and factorization scales, $\alpha_{s}$, and the parton
distribution functions.
The cross sections for \hyperk\ production used in this paper are
computed at leading order only. The theoretical uncertainty
was not evaluated because of the lack of corresponding theoretical
NLO calculations.
For a fixed \hyperrho\ mass, the \hyperk\hyperkbar\ cross section is a
combination of a \hyperrho\ resonance and Drell-Yan production.  When
the \hyperk\ mass is much smaller than  half the \hyperrho\ mass,
Drell-Yan production dominates.  As the \hyperk\ mass increases,
resonance production becomes increasingly important, and dominates as
the kinematic limit for \hyperk\hyperkbar\ pair production is
approached. For \hyperk\
mass greater than  half the \hyperrho\ mass, resonance production
turns off, resulting in a steep drop in the total cross section
(shown by the nearly vertical line in Fig.~\ref{fig::DiChampMassExclusion}).
In addition, near the kinematic limit the \hyperrho $\to$ \hyperk\hyperkbar\
process  produces very low velocity \hyperk\ particles. The signal acceptance
therefore decreases dramatically until the resonance production turns
off, at which point the acceptance increases again. This results in a
spike in the cross section limit near the kinematic limit.

From the intersection of the cross section limit curve
and the central value of the theoretical cross section band, a 95\% CL
lower limit of 1098 (1046)\GeVcc on the mass of
pair produced \PSg, hadronizing into stable $R$-gluonballs
with 0.1 (0.5) probability, is determined with the tracker-only
selection.  The tracker+TOF selection gives a lower limit of 1082 (1030)
\GeVcc for the same signal model.
The analogous limit on the \stone\ mass is 714\GeVcc with the
tracker-only selection and 737\GeVcc with the tracker+TOF
selection.  The charge suppression scenario discussed above yields a
\PSg\ mass limit of  928\GeVcc for $f=0.1$ and 626\GeVcc
for the \stone.
The limits on GMSB and pair produced \stau\ are calculated at 314 and 223\GeVcc, respectively, with the tracker+TOF selection.
The mass limits on \hyperk\ are established at 484, 602 and 747\GeVcc for
\hyperrho\ masses of 800, 1200 and 1600\GeVcc, respectively, with the tracker+TOF selection.

\section{Summary}
The CMS detector has been used to search for highly ionizing,
high-$\pt$ and long time-of-flight massive particles.
Two inclusive searches have been conducted: one that uses highly
ionizing tracks reconstructed in the inner tracker, and a second
requiring that these tracks also be identified in the CMS muon system
and have long time-of-flight. The former is model-independent in that
it is insensitive to the details of $R$-hadron nuclear interactions.
In each case, the observed number of candidates is
consistent with the expected background.
Upper limits on production cross section and lower limits on masses of
stable, weakly- and strongly-interacting particles have been
established for a variety of models. They range from 223\GeVcc for a
pair produced scalar tau to 1098\GeVcc for a pair-produced
gluino. These limits are the most restrictive to date.

\section{Acknowledgements}
We congratulate our colleagues in the CERN accelerator departments for
the excellent performance of the LHC machine. We thank the technical
and administrative staff at CERN and other CMS institutes, and
acknowledge support from: FMSR (Austria); FNRS and FWO (Belgium);
CNPq, CAPES, FAPERJ, and FAPESP (Brazil); MES (Bulgaria); CERN; CAS,
MoST, and NSFC (China); COLCIENCIAS (Colombia); MSES (Croatia); RPF
(Cyprus); MoER, SF0690030s09 and ERDF (Estonia); Academy of Finland,
MEC, and HIP (Finland); CEA and CNRS/IN2P3 (France); BMBF, DFG, and
HGF (Germany); GSRT (Greece); OTKA and NKTH (Hungary); DAE and DST
(India); IPM (Iran); SFI (Ireland); INFN (Italy); NRF and WCU (Korea);
LAS (Lithuania); CINVESTAV, CONACYT, SEP, and UASLP-FAI (Mexico); MSI
(New Zealand); PAEC (Pakistan); MSHE and NSC (Poland); FCT (Portugal);
JINR (Armenia, Belarus, Georgia, Ukraine, Uzbekistan); MON, RosAtom,
RAS and RFBR (Russia); MSTD (Serbia); MICINN and CPAN (Spain); Swiss
Funding Agencies (Switzerland); NSC (Taipei); TUBITAK and TAEK
(Turkey); STFC (United Kingdom); DOE and NSF (USA).

Individuals have received support from the Marie-Curie programme and
the European Research Council (European Union); the Leventis
Foundation; the A. P. Sloan Foundation; the Alexander von Humboldt
Foundation; the Belgian Federal Science Policy Office; the Fonds pour
la Formation \`a la Recherche dans l'Industrie et dans l'Agriculture
(FRIA-Belgium); the Agentschap voor Innovatie door Wetenschap en
Technologie (IWT-Belgium); the Council of Science and Industrial
Research, India; and the HOMING PLUS programme of Foundation for
Polish Science, cofinanced from European Union, Regional Development
Fund.

\bibliography{auto_generated}   

\cleardoublepage \appendix\section{The CMS Collaboration \label{app:collab}}\begin{sloppypar}\hyphenpenalty=5000\widowpenalty=500\clubpenalty=5000\textbf{Yerevan Physics Institute,  Yerevan,  Armenia}\\*[0pt]
S.~Chatrchyan, V.~Khachatryan, A.M.~Sirunyan, A.~Tumasyan
\vskip\cmsinstskip
\textbf{Institut f\"{u}r Hochenergiephysik der OeAW,  Wien,  Austria}\\*[0pt]
W.~Adam, T.~Bergauer, M.~Dragicevic, J.~Er\"{o}, C.~Fabjan, M.~Friedl, R.~Fr\"{u}hwirth, V.M.~Ghete, J.~Hammer\cmsAuthorMark{1}, N.~H\"{o}rmann, J.~Hrubec, M.~Jeitler, W.~Kiesenhofer, V.~Kn\"{u}nz, M.~Krammer, D.~Liko, I.~Mikulec, M.~Pernicka$^{\textrm{\dag}}$, B.~Rahbaran, C.~Rohringer, H.~Rohringer, R.~Sch\"{o}fbeck, J.~Strauss, A.~Taurok, F.~Teischinger, P.~Wagner, W.~Waltenberger, G.~Walzel, E.~Widl, C.-E.~Wulz
% ...
\vskip\cmsinstskip
\dag:~Deceased\\
1:~~Also at CERN, European Organization for Nuclear Research, Geneva, Switzerland\\
% ...

\end{sloppypar}
\end{document}